# Objective crystallographic symmetry classifications of a noisy crystal pattern with strong Fedorov type pseudosymmetries and its optimal image-quality enhancement

Peter Moeck, Department of Physics, Portland State University, Portland 97201-0751, USA, pmoeck@pdx.edu

**Synopsis** Information-theoretic crystallographic symmetry classifications distinguish between genuine symmetries and strong Fedorov type pseudosymmetries in noisy crystal patterns in two dimensions. Because these classifications require neither visual comparisons of image pairs nor subjective interpretations of "symmetry deviation quantifiers" by human beings, they enable the optimal crystallographic processing of an experimental image that results in a significantly enhanced signal to noise ratio of a microscopic study of a crystal.

## Abstract

Statistically sound crystallographic symmetry classifications are obtained with information theory based methods in the presence of approximately Gaussian distributed noise. A set of three synthetic patterns with strong Fedorov type pseudosymmetries and varying amounts of noise serve as examples. Contrary to traditional crystallographic symmetry classifications with an image processing program such as CRISP, the classification process does not need to be supervised by a human being and is free of any subjectively set thresholds in the geometric model selection process. This enables crystallographic symmetry classification of digital images that are more or less periodic in two dimensions (2D), a.k.a. crystal patterns, as recorded with sufficient structural resolution from a wide range of crystalline samples with different types of scanning probe and transmission electron microscopes. Correct symmetry classifications enable the optimal crystallographic processing of such images. That processing consists in the averaging over all asymmetric units in all unit cells in the selected image area and significantly enhances both the signal to noise ratio and the structural resolution of a microscopic study of a crystal. For sufficiently complex crystal patterns, the information-theoretic symmetry classification methods are more accurate than both visual classifications by human experts and the recommendations of one of the popular crystallographic image processing programs of electron crystallography.

## 1. Introduction: the paper's background, organization, motivation, primary goal and secondary objective

### 1.1. Crystallographic symmetries and pseudosymmetries

The symmetries of the Euclidean plane that are compatible with translation periodicity in two dimensions (2D) are tabulated exhaustively in volume A of the International Tables for Crystallography (Aroyo, 2016) and in the brief teaching edition (Hahn, 2010) of that series of authoritative reference books from the International Union of Crystallography (IUCr). Noncrystallographic symmetry has been defined in the on-line dictionary of the IUCr as a *"symmetry operation that is not compatible with the periodicity of a crystal pattern"* (Dictionary I).

It is also noted in this dictionary and by Nespolo and co-workers (2008) that this term is often improperly used in biological crystallography, where one should refer either to local and partial symmetry operations, on the one hand, and pseudosymmetries, on the other hand. The above-mentioned on-line directory defines a crystallographic pseudosymmetry simply as featuring a *"deviation"* from a space group symmetry (of one, two, or three dimensions) that *"is limited"* without explaining how the deviation is to be quantified (Dictionary II). In this paper, we will provide such quantifications for three synthetic crystal patterns.

A crystal pattern is defined as the *"generalization of a crystal structure to any pattern, concrete or abstract, in any dimension, which obeys the conditions of periodicity and discreteness"* (Dictionary III). Physical realizations of crystal pattern can be undisturbed or disturbed/noisy.

Pseudosymmetry is *"a spatial arrangement that feigns a symmetry without fulfilling it"* (Moeck, 2018) and can exist in direct space at either the site/point symmetry level of a plane symmetry group or the projected Bravais lattice type level, or a combination thereof. When a very strong translational pseudosymmetry results in metric tensor components and lattice parameters that are, within experimental error bars, indistinguishable from those of a higher symmetric Bravais lattice type, one speaks of a metric specialization (Moeck and DeStefano, 2018). On the site/point symmetry level, one can make a distinction between crystallographic pseudosymmetries that are either compatible with the Bravais lattice of the unit cell of the genuine symmetries or a sublattice of the genuine symmetries. These kinds of pseudosymmetries are often collectively called Fedorov type pseudosymmetries (Chuprunov, 2007).

Pseudosymmetries of the Fedorov type form plane "pseudosymmetry groups", which are either disjoint or non-disjoint from the plane symmetry group and projected Laue class of the genuine symmetries. The lowest symmetric pseudosymmetry groups is per definition always disjoint from the lowest symmetric genuine symmetry group that provides the best fit to experimental data. The minimal Fedorov type pseudosymmetry supergroups of lowest symmetric maximal pseudosymmetry subgroups can, however, be non-disjoint from the lowest symmetric genuine symmetry group.

When Fedorov type pseudosymmetries and genuine symmetries exist in direct space, they exist in reciprocal/Fourier space as well. In noisy experimental data, local and partial symmetries may become difficult to distinguish from pseudosymmetries and genuine symmetries alike.

## 1.2. Assignments of symmetries in the presence of noise

Note that only the idealized structure of a real-world crystal is strictly periodic in three dimensions (3D) and features an unbroken discrete space symmetry group. Analogously, the idealized structure of a subperiodic crystal (such as a regular array of intrinsic membrane protein complexes in a lipid bilayer) is strictly periodic in 2D and features an unbroken discrete layer symmetry group (Kopský and Litvin, 2010).

The 2D projection of the structure of a *real* crystal that contains only a few localized symmetry breaking structural defects is, however, deemed to possess a discrete plane symmetry group on average over multiple unit cells as well. The genuine plane symmetry group of the projected real crystal structure is per definition the plane symmetry group that is least broken. The lowest symmetric plane symmetry group of the genuine symmetries is referred here to as the "anchoring group" and is measurably least broken in the crystal pattern by "aggregated noise" from multiple sources.

By these definitions, Fedorov type pseudosymmetry groups are broken to a measurably larger extent than the symmetry group of the genuine symmetries (and all maximal subgroups of these symmetries and their respective maximal subgroups). This will be further elaborated on in the second section of this paper, where a visual example is provided.

In the presence of noise, it may become difficult for human classifiers to distinguish Fedorov type pseudosymmetries from their genuine symmetries counterparts. This difficulty arises from the unaided human classifier's need to extrapolate "on sight" to a hypothetical noise-free version of the crystal pattern.

## 1.3. Crystallographic image processing and the symmetry inclusion problem

The essence of crystallographic image processing (Hovmöller, 1992, Valpuesta and co-workers, 1994, Wan and co-workers, 2003, Kilaas and co-workers, 2005, Gipson and co-workers, 2007, Zou and co-workers, 2011) is the enforcing of the 2D site/point symmetries that correspond to a certain higher symmetric plane symmetry group on all of the pixel intensity values within the direct-space translation-averaged unit cell.

The Fourier space representation of the translation-averaged unit cell is obtained by calculating the discrete Fourier transform of the image intensity and the filtering out of all non-structure-bearing Fourier coefficients. The Fourier-back transforming of the periodic structure-bearing Fourier coefficients (that are laid out on a reciprocal lattice in the amplitude map of the discrete Fourier transform) leads to the translation-averaged unit cell in direct space.

Obtaining the translation-averaged direct space unit cell is, therefore, known as traditional Fourier filtering (Park and Quate, 1987). The non-structure bearing Fourier coefficients represent the bulk of the noise in the direct space image. Accordingly, their filtering out enhances the signal to noise ratio and structural resolution of the Fourier-filtered image.

The enforcing of the symmetries of a certain higher symmetric plane symmetry group on the structure-bearing Fourier coefficients of a more or less 2D periodic image is loosely speaking obtained by averaging over the corresponding symmetry related sets of structure-bearing Fourier coefficients. (These sets are specific to each plane symmetry group.) This averaging/symmetrizing enforces all site/point symmetries of the chosen plane symmetry group onto the translation-averaged unit cell when the symmetrized structure-bearing Fourier coefficients are back-transformed into a direct space image. In effect, one has averaged in Fourier space over all asymmetric units in all unit cells of a selected region of a digital direct-space input image.

When done correctly, crystallographic image processing increases the signal to noise ratio and intrinsic quality[1] (Paganin and co-workers, 2019, Gureyev and co-workers, 2019) of a digital image in direct space significantly. Compared to traditional Fourier filtering, the processing of a digital image in the correctly determined plane symmetry group leads to a further increase of the signal to noise ratio and an associated increase of the structural resolution of a crystallographic study. For (approximately) Gaussian distributed noise, crystallographic image processing is by (approximately) the square root of the multiplicity of the general position per lattice point more effective in the suppression of noise than Fourier filtering alone. (That multiplicity is equal to the number of non-translational symmetry operations in a plane symmetry group.)

---

[1] Crystallographic image processing is in an appendix to arXiv: 2108.00829, i.e. the expanded version of this paper, discussed as a form of computational imaging. The concept of intrinsic image quality is defined there by means of an equation. The concept of "Abbe resolution" is also defined in the main part of that open-access paper.

The knowledge of the most likely plane symmetry that a hypothetical version of an image would possess in the absence of noise is the precondition for the correct/optimal crystallographic processing of that image. For a previously not classified crystal or crystal pattern, this knowledge has historically not been easy to come by. Elucidating that kind of plane symmetry group has been a long-standing problem in both the computational symmetry subfield of computer science (Liu and co-workers, 2009) and electron crystallography.

The main reason that this problem had remained unsolved for more than half a century is the existence of mathematically defined inclusion relations between the individual crystallographic symmetry groups, classes, and types. In other words, the main reason was the non-disjointness of many of the geometric models that are to be compared to the input image data and from which the best, i.e. statistically most justified, model for the digital input image data is to be selected. Symmetry inclusion relations, non-disjointness, and disjointness are explained in some detail in the third section of this paper. That section also presents the plane symmetry hierarchy tree as a visualization of disjoint and non-disjoint symmetry inclusion relationships between the translationengleiche (Aroyo, 2016, Hahn, 2010, Burzlaff and co-workers, 1968) maximal subgroups and minimal supergroups of the plane symmetry groups. The symmetry hierarchy tree of the 2D point symmetries that are projected Laue classes is also provided there.

## 1.4. Using a geometric form of information theory[2] offers a workaround to the symmetry inclusion problem

This author presented recently so far unique interpretation-threshold-free solutions to identifying the genuine plane symmetry group and projected Laue class in digital more or less 2D periodic images in the presence of pseudosymmetries and generalized noise (Moeck, 2018 and 2019, Moeck and Dempsey, 2019, Dempsey and Moeck, 2020, Moeck, 2021). Fedorov type pseudosymmetries do not present challenges to these solutions as they are reliably identified (and can be quantified) as long as noise levels are moderate. This will be demonstrated in this paper.

The author's solutions are based on Kenichi Kanatani's geometric form of information theory[2] (Kanatani, 1997, 1998, 2004, and 2005). Kanatani's theory presents a geometric "workaround" to the symmetry inclusion relations problem and has the added benefit that the prevailing noise level does not need to be estimated for the comparison of non-disjoint geometric models of digital image data. This statistical theory tackles the inclusion problem that a less restricted, e.g. lower symmetric, model of some input image data will always feature a smaller deviation (by any kind of distance measure) to the input image data than any more restricted, e.g. higher symmetric, model that is non-disjoint (Kanatani, 1997 and 1998). In other words, the fit to some experimental data with more parameters will always be better than a fit with fewer parameters. The adaptation of Kanatani's framework to crystallographic symmetry classifications and quantifications is described in detail in Moeck (2018). The third section of this paper gives the relevant equations and inequalities for making objective plane symmetry and projected Laue class classifications with the author's methods. (The usage of those relations has led to the results that are presented in the fourth section.)

Objectivity is in this paper to be understood as only stating what digital image data actually reveal about a crystallographic symmetry without any subjective interpretation of any symmetry distance measure. This objectivity is obtained by using a geometric form of information theory.

Note that the information theory based crystallographic symmetry classification methods of this author should be generalized to three spatial dimensions. This is because there is also subjectivity in the current practice of single crystal X-ray and neutron crystallography (Moeck, 2018). Fedorov type pseudosymmetries exist also in three dimensions and are not rare in nature (Chuprunov, 2007, Somov and Chuprunov, 2009, Moeck 2018). The symmetry inclusion relationships of the space groups occupy the bulk of volume A1 of the International Tables for Crystallography (Wondratschek and Müller, 2004). Note in passing that Kanatani's statistical theory is valid in any dimension.

It is very well known that the structural resolution of crystallographic studies depends on the number of structural entities over which one averages (McLachlan, 1958). The optimal averaging can, however, only be obtained for the correct prior symmetry classification of the data that enter into such studies when no prior knowledge of the crystal and/or crystal pattern symmetry is available.

Optimal crystallographic averaging in 2D and crystallographic image processing on the basis of the correctly identified plane symmetry group are synonymous. One enforces in this case all of the site/point symmetries that the translation averaged unit cell image needs to feature in order to be the best representation of the input image data in the information-theoretic sense. This best representation is often called the "Kullback-Leibler best", "minimal geometric Akaike Information Criterion (G-AIC) value", or simply, K-L best geometric model that the input image data maximally supports.

---

[2] According to the Merriam-Webster Dictionary, information theory is defined as *"a theory that deals statistically with information, with the measurement of its content in terms of its distinguishing essential characteristics or by the number of alternatives from which it makes a choice possible, and with the efficiency of processes of communication between humans and machines"*. https://www.merriam-webster.com/dictionary/information theory.

## 1.5. Prior information-theoretic distinctions between genuine symmetries and Fedorov type pseudosymmetries based on a reasonable noise distribution estimate

Generalized noise (Moeck, 2018 and 2019, Dempsey and Moeck, 2020) is defined in this paper as the sum of all deviations from the genuine translation periodic symmetries in a crystal's structure and/or the imaged 2D periodic properties of the crystal. At the experimental level, generalized noise as defined here combines all effects of a less-than-perfect imaging of a crystal, all rounding errors and effects of approximations in the applied image processing algorithms, effects such as uneven staining in the cryo-electron microscopy of subperiodic intrinsic membrane protein crystals, slight deviations from exact zone-axis orientations in transmission electron microscopes, and the real structure that typically exist in addition to the ideal structure of a crystal. This definition applies also to undisturbed and disturbed/noisy crystal patterns in two dimensions as analyzed in this paper. For the author's information-theoretic crystallographic symmetry classification methods (Moeck, 2018 and 2019, Moeck and Dempsey, 2019, Dempsey and Moeck, 2020, Moeck, 2021) to work reliably, the generalized noise needs to be Gaussian distributed (with mean zero and standard deviation $\varepsilon$, which Kanatani calls the "*noise level*", Kanatani, 2005) to a sufficient approximation.

The information-theoretic distinction between Fedorov type pseudosymmetries that are compatible with a sublattice of the underlying Bravais lattice from the genuine symmetries has been demonstrated already in a very short conference paper (Moeck and Dempsey, 2019). Those symmetry classifications used a crystal pattern of low complexity to which moderate to large amounts of Gaussian distributed noise were added.

Dempsey and Moeck (2020) simulated the amounts and types of noise that needed to be added to a crystal pattern with site/point and translational pseudosymmetries for the plane symmetry classifications by the information theoretic method to misclassify pseudosymmetries as genuine symmetries. Fourteen versions of the same medium complexity pattern were used in that study. For each version, four classifications were made for pattern regions of different sizes and shapes. The addition of strictly Gaussian distributed noise, up to the limit that a freely available computer program (Program I) enabled, did not result in any misclassification. Changing the aggregate composition of the noise systematically so that it was to lesser extents approximately Gaussian distributed resulted in a single misclassification (out of 56 classifications in total.) The misclassification happened for the noisiest image and the smallest image-region selection. Note that human expert classifiers would probably have made more than one misclassification when confronted with the same tasks (Dempsey and Moeck, 2020).

As it is time to, this paper will demonstrate statistically sound distinctions between genuine symmetries and strong Fedorov type pseudosymmetries for a highly complex crystal pattern and two of its noisy versions in its fourth section.

## 1.6. Crystallographic symmetry classifications and image processing in contemporary electron crystallography

The common practice in electron crystallography is to make crystallographic symmetry classification on the basis of subjective interpretations of the values of Fourier-space "symmetry deviation quantifiers" that measure distances between the translation-averaged input image and differently symmetrized versions of that image (Hovmöller, 1992, Zou and co-workers, 2011, Gipson and co-workers, 2007, Wan and co-workers, 2003, Kilaas and co-workers, 2005, Henderson and co-workers, 2012, Lawson and co-workers, 2020). Following up on a report by Henderson and co-workers (2012) on the first electron crystallography validation task force meeting, it has recently been noted with respect to cryo-electron microscopy that *"… as currently practiced, the procedure is not sufficiently standardized: a number of different variables (e.g. … threshold value for interpretation) can substantially impact the outcome. As a result, different expert practitioners can arrive at different resolution estimates for the same level of map details."* (Lawson and co-workers, 2020). In the context of computational imaging[1] (Gureyev and co-workers, 2019, Paganin and co-workers, 2019), "resolution" in this direct quote stands for structural resolution and intrinsic image quality.

Two different sets of structure-bearing Fourier coefficient based symmetry deviation quantifiers, as implemented in the crystallographic image processing programs CRISP (Hovmöller, 1992, Zou and co-workers, 2011, Zou and Hovmöller, 2012) and ALLSPACE (Valpuesta and co-workers, 1994), are most popular in the electron crystallography community. Neither of these two sets of quantifiers are maximal likelihood estimates combined with geometric-model selection-bias correction-terms for objective symmetry model selections of digital input image data. A geometric form of information theory can, therefore, not be based on these quantifiers in order to avoid a necessarily subjective decision of what the underlying plane symmetry most likely is (in the considered opinion of the users of these two computer programs).

Whereas the sets of typically employed symmetry deviation quantifiers in contemporary electron crystallography provide quantitative numerical measures, the decision as to which plane symmetry group should be enforced on the input image data as part of its crystallographic image processing is with necessity left to the electron crystallographer. In the presence of symmetry inclusion relations, Fedorov type pseudosymmetries, and generalized noise, optimizing the fit between geometric models for experimental data and the data itself by minimizing symmetry deviation quantifiers and using overriding rules of thumb such as "when in doubt, choose the higher symmetry" (Hovmöller, 2010, Zou and co-workers, 2011, Zou and Hovmöller, 2012, Eades 2012) are certainly not a foolproof strategy for optimal model selection.

The CRISP program makes a suggestion that the user may either accept or overwrite, but relies heavily on visual comparisons between differently symmetrized versions of the input image data. This author has not used ALLSPACE (in its *2dx*, Gipson and co-workers, 2007, and *Focus*, Biyani and co-workers, 2017, incarnations) so far as no version that runs on Microsoft Windows compatible computers seems to exist. There are also competing computer programs with less comprehensive symmetry deviation quantifiers, e.g. VEC (Wan and co-workers, 2003) and EDM (Kilaas and co-workers, 2005) that rely even more heavily on visual comparisons of the translation-averaged image to its symmetrized versions.

When the underlying plane symmetry in a noisy experimental image has been underestimated, i.e. only a subgroup of the most likely plane symmetry group has been identified, one does not make the most out of the available image data in the subsequent symmetry enforcing step of the crystallographic image processing procedure. On the other hand, if the plane symmetry is overestimated, "non-information" due to noise will unavoidably be averaged with genuine structural information in the subsequent crystallographic processing of the image. In the latter case, one may have wrongly identified a minimal supergroup of the correct plane symmetry group that the analyzed image would possess in the absence of generalized noise. That supergroup could be the union of a genuine plane symmetry group and a Fedorov type pseudosymmetry group.

It is, accordingly, very important to get the crystallographic symmetry classification step of the crystallographic image processing procedure just right. For that, one should only rely on the digital image data itself and refrain from any subjective considerations.

With the author's objective and interpretation-threshold-free methods (Moeck, 2018 and 2019, Moeck and Dempsey, 2019, Dempsey and Moeck, 2020, Moeck, 2021), one can now make advances with respect to the above stated situation in the cryo-electron microscopy subfield that deals with subperiodic intrinsic membrane protein crystals, in the electron crystallography of inorganic materials, and the crystallographic processing of digital crystal patterns in general.

## 1.7. Primary goal and secondary objective of this paper

The primary goal of this paper is to demonstrate the author's interpretation-threshold-free crystallographic symmetry classification methods on a series of three synthetic crystal patterns, where one is free of noise and the other two are noisy. The achievement of this goal might entice the computational symmetry and electron crystallography communities to replace their subjectivity in crystallographic symmetry classifications with the objectivity that the information theory based methodology enables.

The demonstration of the benefits of the correct crystallographic processing of a more or less 2D periodic image is the secondary objective of this paper. Scanning probe microscopists should take notice as these demonstrations are mainly directed to them. This is because crystallographic image processing is just as applicable to more or less 2D periodic images from scanning probe microscopes (Moeck, 2017 and 2021) as it is to images from parallel-illumination transmission electron microscopes (as used in electron crystallography).

Scanning probe microscopists may, however, like to correct for scanning distortions in their images of 2D periodic samples with tools such as Jitterbug (Jones and Nellist, 2013) before they make crystallographic symmetry classifications and process their images crystallographically. The achievement of the secondary objective, i.e. demonstrating the benefits of the correct crystallographic processing of a more or less 2D periodic image, may eventually lead to the widespread use of crystallographic image processing techniques in scanning probe microscopy.

The limiting effects of noise and Fedorov type pseudosymmetries in more or less 2D periodic images on the accuracy of crystallographic symmetry classifications have so far rarely been analyzed. As one would expect, the distinction between genuine symmetries and pseudosymmetries of the Fedorov type becomes more difficult with increasing amounts of noise even when a geometric form of information theory is used (Moeck and Dempsey, 2019, Dempsey and Moeck, 2020). This will be demonstrated here once more in the fourth section of this paper. That section constitutes this paper's main part and features four subsections containing nine numerical data tables as well as four figures. Two of these figures demonstrate the beneficial noise reduction and crystallographic-averaging-induced structural resolution enhancement effects of crystallographic image processing.

In order to facilitate direct comparisons to results obtained by one of the two most popular traditional crystallographic symmetry classification programs of electron crystallography, *.hka files were exported from the CRISP program and used for the calculation of the ratios of sums of squared residuals of non-disjoint geometric models for the image input data.

The fifth section of this paper compares the results of our three crystallographic symmetry classifications (by the author's information theory based methods) to plane symmetry group estimates by the program CRISP as applied to the same and adjacent areas of the three synthetic crystal patterns. The paper ends with a summary and conclusions section.

## 1.8. The three appendices of this paper

Appendix A provides "Notes to the text". They are in essence expanded footnotes. Analogously to footnotes, they are in the main text marked by superscripts[Ax] on a key word, where x is an integer starting with unity. For example, a brief account of the physical creation[A1] of the undisturbed crystal pattern that is analyzed in this paper is given in that appendix as note A1, as it is the first of such notes. From the account in that particular end-note, it is obvious that the accurate symmetry classification of the crystal pattern in Figure 1 can only be plane symmetry group *p4*. Strong pseudosymmetries of the Fedorov type are present in this pattern that human classifiers will, at least at first sight, most likely misinterpret as the genuine symmetries of plane symmetry group *p4gm*.

Appendix B presents the formulae for ad-hoc defined confidence levels for classifications into minimal supergroups of the genuine symmetries for the special case that all geometric models of the digital input image data are based on the same number of structure-bearing Fourier coefficients. Outlooks on ongoing developments of the information theory based crystallographic symmetry classification and quantification methods and some of their potential applications[A2] are provided in appendix C.

## 2. Fedorov type pseudosymmetries illustrated on a noise-free synthetic pattern

Figure 1 shows a slightly enlarged reproduction of a crystal pattern that originated with the artist Eva Knoll (Knoll, 2003). There are about 15.5 translation periodic motifs in the digital representation of this particular graphic work of art in Knoll's paper.

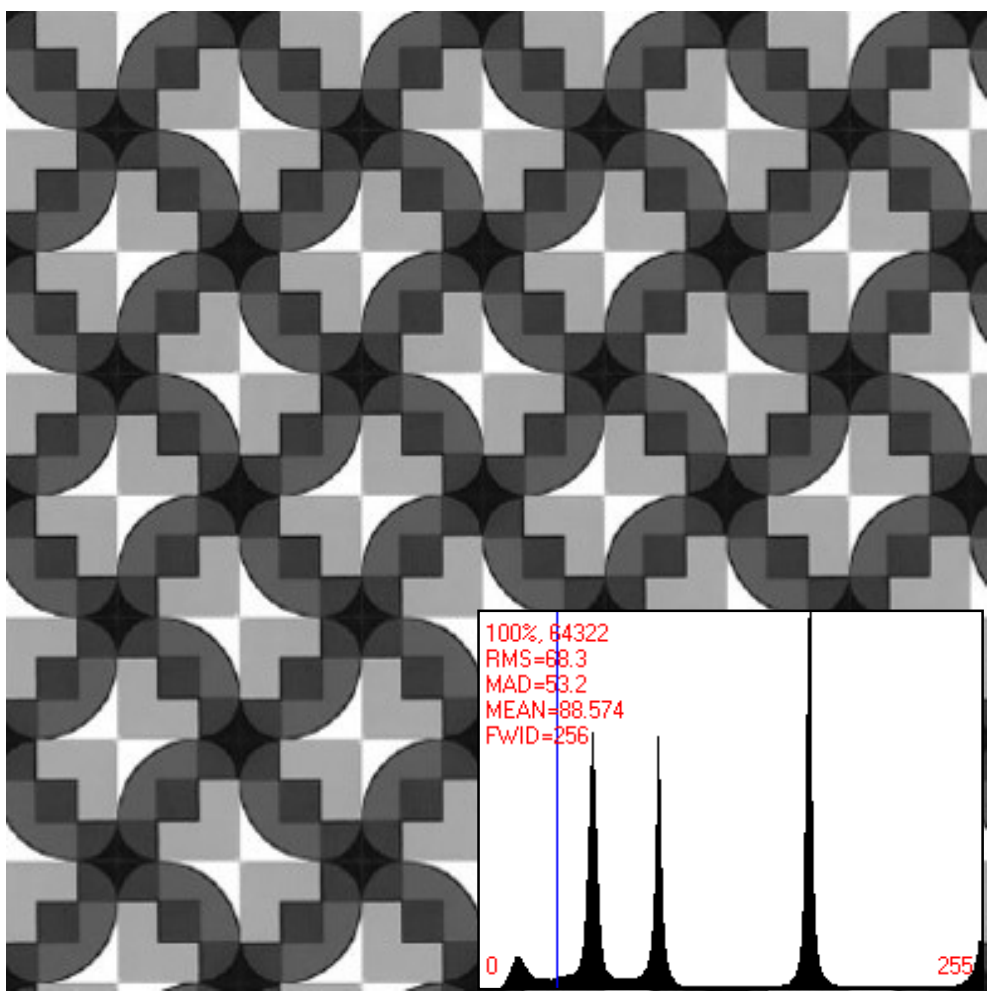

**Figure 1.** Section of an expanded digital version of the graphic artwork "Tiles with quasi-ellipses" (1992, acrylic on ceramic) by Eva Knoll.
Histogram of the whole crystal pattern as inset. The vertical thin line and descriptive annotations in the histogram are due to the computer program CRISP.
Note for references below the "bright bow tie" feature with a pixel intensity of around 255, and the "dark curved diamond" feature with an intensity level of around 21. The histogram entries are explained in the expanded on-line version of this paper, arXiv: 2108.00829.

After expansion by periodic motif stitching of a digital representation of the original artwork as presented in Knoll (2003), that pattern featured approximately 144 primitive unit cells in total. Approximately 16 of these unit cells are shown in Figure 1. The computer program Image Composite Editor (Program II) was used for the periodic motif stitching. The expanded image/crystal pattern is provided in the supporting material of this paper in the *.jpg format (1160 by 1165 pixels with 24 bit depth, and 413.058 bytes) as well as in the uncompressed *.tif format (1160 by 1165 pixels with 32 bit depth, 120 by 120 dpi, resolution unit 2, color representation sRGB, attribute A, and 5.442.642 bytes). Just as in Dempsey and Moeck (2020), the periodic motif stitching was done in order to enable more precise crystallographic analyses.

The stitched/expanded crystal pattern (of which Figure 1 shows a small section) serves in this paper as the basis of three synthetic patterns that are to be classified with respect to their crystallographic symmetries and Fedorov type pseudosymmetries. The two per design noisy versions of the crystal pattern (in the series of analyzed patterns) are processed crystallographically in order to demonstrate that technique's benefits with respect to the noise suppression and site/point symmetry enforcing of such a processing.

Because the physical piece of graphic art from which the digital pattern in Figure 1 was created is hand made[A1], none of the 2D translation compatible crystallographic symmetries of the Euclidean plane are strictly speaking present as they are only mathematical abstractions. It is, however, standard practice to assign a plane symmetry group to such a crystal pattern as one would also do for any sufficiently well resolved image from a real crystal in the real world, see section 1.2 above. That symmetry group of the pattern or image is per definition the one that is least broken by structural, sample preparation, imaging, and image processing imperfections (generalized noise).

For the purpose of the crystallographic symmetry classification, the assumption is made that the imaging and image processing imperfections of the crystal pattern in Figure 1 are negligible and that there are no structural imperfections/defects that are intrinsic to the represented physical object. The generalized noise in that pattern is, therefore, negligible and we call the corresponding pattern the noise-free member of a series of three crystal patterns that are to be classified with respect to their crystallographic symmetries and Fedorov type pseudosymmetries in this paper.

A human expert classifier would most likely assign plane symmetry group *p4gm* to the crystal pattern in Figure 1 at first sight because approximate four-fold and two-fold rotation points as well as mirror and glide lines are all visibly recognizable in their required spatial arrangements in all of the 2D translation periodic unit cells. (This author assigned plane symmetry group *p4gm* to the pattern in this figure as well at first sight, but corrected his mistake after a more careful visual analysis.)

The different types of visually recognizable point/site symmetries in each individual unit cell are probably broken by slightly different amounts, but these differences appear to be so minor that a human being may just assume they are all broken by the same amount. Under this assumption, plane symmetry group *p4gm* would indeed underlie the completely symmetric idealization of the crystal pattern in Figure 1. The rather sharp peaks in the histogram in Figure 1 are to be interpreted as genuine characteristics of the underlying crystal pattern since no noise was added to deliberately disturb this pattern.

The image-pixel-value based classification of this crystal pattern with the author's method reveals, however, plane symmetry groups *p2* and *p4* as genuine, with *p2* least broken being the anchoring group, and the Fedorov type pseudosymmetry groups *p1g1*, *p11g*, *c1m1*, and *c11m* as quantitatively more severely broken than the *p2* and *p4* symmetries. These pseudosymmetries combine with the genuine symmetries to the two minimal pseudosupergroups *p2gg* and *c2mm*, as well as their respective minimal pseudosupergroup *p4gm*. (With hindsight, this is as it must be given the sequence of creative processes[A1] that resulted in this particular graphic piece of art.) The fourth section of this paper gives the details of the corresponding analysis.

The point/site symmetry of the centers of the conspicuous bright "bow ties" in this pattern is visibly no higher than point symmetry group *2*, which is one of the maximal subgroups of *2mm*. Site symmetry *2mm* is, on the other hand, one of the minimal supergroups of point symmetry group *2*, but visibly more severely broken in the crystal pattern in Figure 1.

This becomes even clearer in Figures 2 and 3. Approximately four primitive (or two centered) unit cells of the pattern in Figure 1 are displayed in Figure 2 after translation averaging by Fourier filtering[A3]. Note that each bright bow tie in Figure 2 is shared between two adjacent unit cells that are based on what seems to be a square Bravais lattice. The centers of the bright bow ties are at fractional unit cell coordinates ½,0, ½,1, 0,½, and 1,½, as marked by insets in Figure 2.

These points feature visually the approximate site symmetry *2* at best, rather than *2mm*, which would be required if the underlying plane symmetry group were to be *c2mm* or *p4gm*. The observed site symmetry *2* at these fractional unit cell coordinates is, on the other hand, compatible with plane symmetry groups *p2*, *p2gg*, and *p4*.

At the fractional unit cell coordinates 0,0, 1,0, 0,1, and 1,1 as well as ½,½ in Figure 2, there are also approximate four-fold rotation points at the centers of dark "curved diamonds" so that a *p4* or *p4gm* classification by a human expert is probably the best anyone could come up with when the slight differences in the breaking of the individual symmetry operations are not noticed and quantified. The genuine plane symmetry group of this pattern can, however, only be *p2*, *p2gg*, or *p4* when the visible site/point symmetry around the centers of the bright bow ties is taken into account.

Figure 3 zooms into the translation periodic motif of Figure 2 and features a single bright bow tie and its immediate surrounding.

Both of the arrows in Figure 3 point to positions in the motif where the tips of the bright bow ties end and meet straight edges from the gray "right angle ruler" parts of the motif. There is approximately a 20 % difference in the distance of these points from the horizontal and vertical edges of the gray right angle ruler shaped motif parts, so that there is definitively no mirror line from the top-right corner to the bottom left corner in this figure. Such a mirror line would be required for the whole motive to be part of a primitive unit cell with plane symmetry group *p4gm* or a centered unit cell with plane symmetry group *c2mm*.

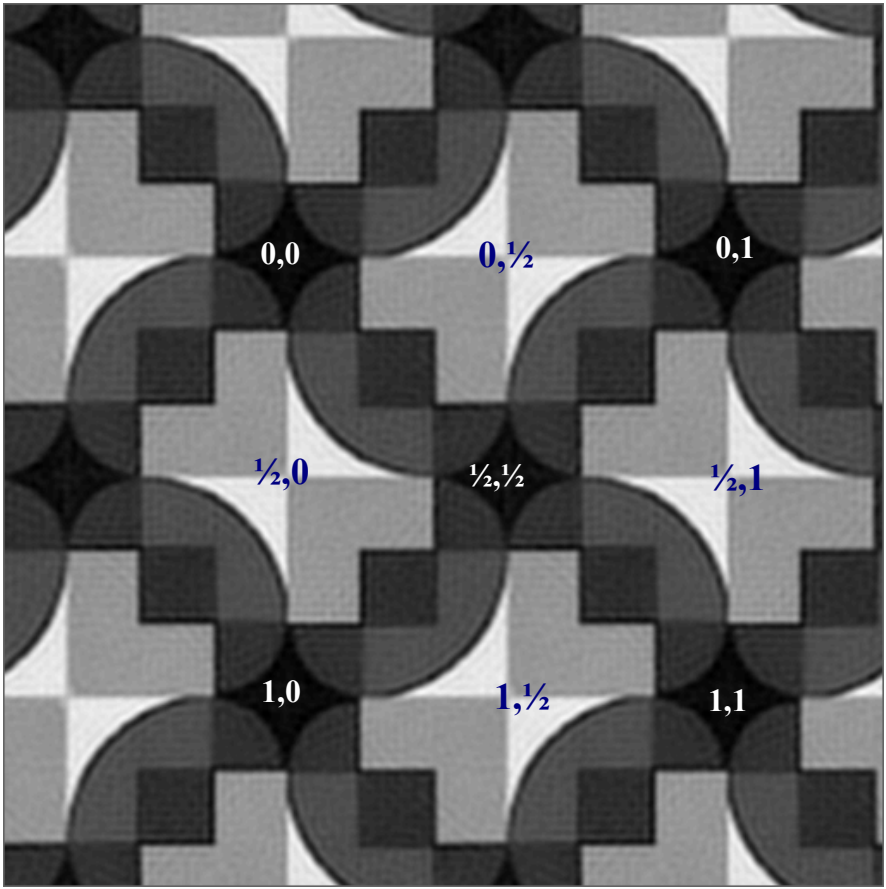

Figure 2. Approximately four primitive (or two centered) translation averaged unit cells of the crystal pattern in Figure 1 after Fourier filtering over approximately 88 stitched-together primitive unit cells and using the strongest 956 structure-bearing Fourier coefficients in the Fourier back-transform to direct space. Selected fractional unit cell coordinates as insets.

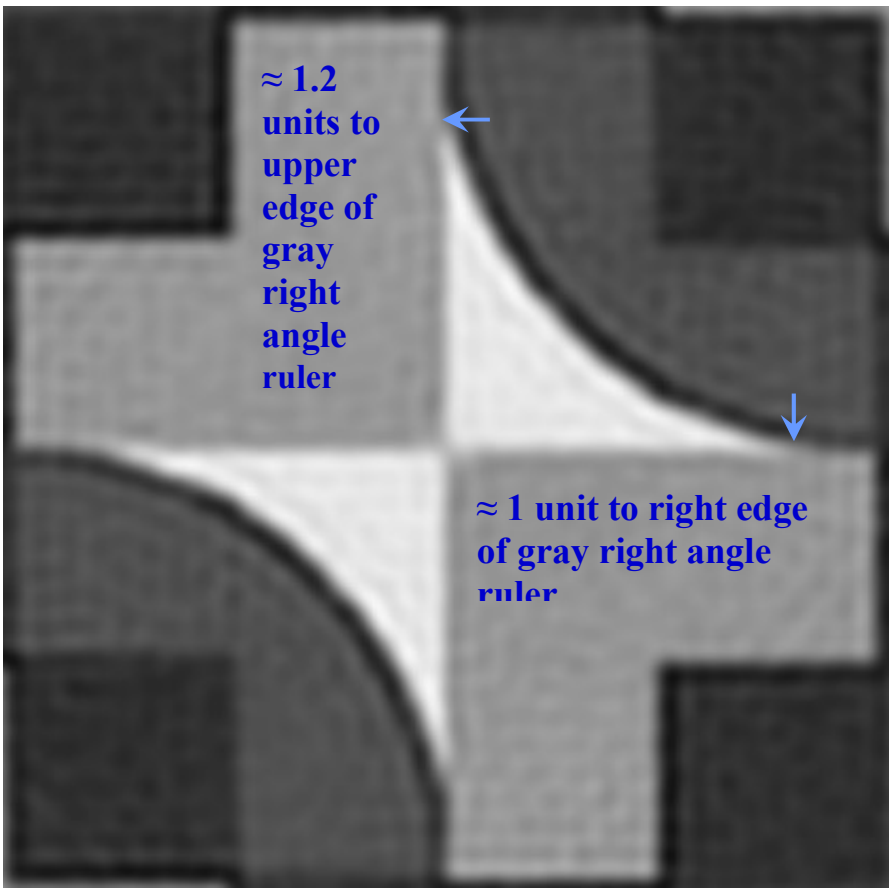

Figure 3. One bright bow tie in a close-up of Figure 2. There is probably no longer an argument that the point symmetry of this feature is at best point group *2*.

## 3. Pertinent equations, inequalities, plane symmetry and 2D Laue class hierarchy trees; and their usages

Kanatani's G-AIC relies on the noise being approximately Gaussian distributed. For that kind of noise, the residuals need to be sums of squares of the differences between the input data and geometric models for that data. Since crystallographic symmetry classifications are best done in Fourier space, the maximal likelihood estimate for approximately Gaussian distributed noise in more or less 2D periodic patterns takes the form of the sums of squared residuals of the complex structure-bearing Fourier coefficients for plane symmetry group classifications. For projected Laue class classifications, they take the form of the sums of squared residuals of the amplitudes of those Fourier coefficients.

Equation (1) gives the sum of squared residuals of the complex Fourier coefficients of a symmetrized (geometric) model of the input image data with respect to the translation-averaged-only (Fourier filtered) version of this data:

$$\widehat{J}_{cFC} = \sum_{j=1}^{N} (F_{j,trans} - F_{j,sym})^* \cdot (F_{j,trans} - F_{j,sym}) \tag{1},$$

where $(.)^*$ stands for the complex conjugate of the difference of a pair of complex numbers $(.)$. The sum is over the differences of all $N$ structure-bearing Fourier coefficients with matching Laue indices, and the subscripts on the right-hand side stand for *translation averaged* and *symmetrized*, respectively. The subscript on the left-hand side stands for *complex Fourier coefficients*. Note that there is a zero sum of residuals per equation (1) for the case of $F_{j,trans} = F_{j,sym}$, i.e. the translation-averaged-only model of the input image data, which features plane symmetry group *p1*.

The sum of squared residuals of the amplitudes of the Fourier coefficients is calculated in an analogous manner from the real valued amplitudes of the structure-bearing Fourier coefficients:

$$\widehat{J}_{aFC} = \sum_{j=1}^{N} (\left| F_{j,trans} \right| - \left| F_{j,sym} \right|)^2 \tag{2}$$

where the subscript on the left-hand side stands for *amplitude of Fourier coefficients*.

Note again that the sum of residuals is zero when all of the translation averaged and symmetrized Fourier coefficient amplitudes with matching Laue indices are equal to each other. This happens for the translation-averaged-only model of the input image data, which features point symmetry group *2* due to the Fourier transform being centrosymmetric. Projected Laue class *2* features, accordingly, a zero sum of amplitude residuals in the data tables that are shown in the fourth section of this paper.

In order to restrict the sums of squared residuals to small numbers, the structure-bearing Fourier coefficients of the input image intensity and their symmetrized versions are in this paper normalized through division by the maximal amplitudes that the CRISP program provides for both the translation averaged model and the symmetrized models of the input image data in both equations (1) and (2).

What follows below is valid for classifications into both plane symmetry groups and projected Laue classes. The same equations and inequalities as well as analogous considerations concerning the plane symmetry group hierarchy and the hierarchy of 2D point groups that are projected Laue classes apply so that the subscripts *cFC* and *aFC* on the sums of squared residuals from equations (1) and (2) are dropped below. Two different symmetry hierarchy trees will, however, be applicable. The first one for plane symmetry groups is presented in Figure 4a below. The second one is given in Figure 4b for projected Laue classes.

Kanatani's G-AIC has the general form:

$$G - AIC(S) = \widehat{J} + 2(dN + n)\widehat{\varepsilon}^2 + O(\widehat{\varepsilon}^4) + ... \tag{3},$$

where $\widehat{J}$ is a sum of squared residuals, as for example given in equations (1) and (2), for the geometric model $S$, $d$ is the dimension of $S$, $N$ is the number of data points that represent the model $S$, $n$ is the number of degrees of freedom of $S$, and $\widehat{\varepsilon}^2$ is the variance of a generalized noise term, which obeys a Gaussian distribution to a sufficient approximation. The $O(\widehat{\varepsilon}^4)$ term in (3) represents unspecified terms that are second-order in $\widehat{\varepsilon}^2$, while the ellipsis indicates higher-order terms that become progressively smaller.

For small and moderate amounts of generalized noise, it is justified to ignore all of the higher-order terms in (3)

$$G - AIC(S) = \widehat{J} + 2(dN + n)\widehat{\varepsilon}^2 \tag{4},$$

because they will make only minor contributions to the G-AIC values of all geometric models. The number of data points, $N$, can either be constant for all geometric models in a set of models or differ from model to model but should in the latter case be on the same order. The dimension of the model is defined by the geometric type of model. (Note in passing that Kanatani refers to the equivalent of (4) as *normalized geometric AIC*, involving *normalized residuals* and *normalized covariance matrices that are isotropic*, in his monograph, and designates it as $AIC_0(S)$, Kanatani 2005).

Equation (4) is to be interpreted as a "balanced geometric model residual" for geometric model selections that is well suited to deal with symmetry inclusion relations. A non-disjoint and less-constrained model, which is lower symmetric, will always fit the input data better than the more constrained model that features a higher non-disjoint symmetry. The $\hat{J}$ value of the less constrained (more general) model that is in a non-disjoint relationship with a higher symmetric model will, therefore, be smaller than its counterpart for the more constrained model. In other words, the more general model fits the data better than the more restricted model. This is because the more general (less constrained) model has more degrees of freedom.

As long as the G-AIC value of a *more* constrained (higher symmetric) model, subscript $m$, is smaller than that of the *less* constrained (less symmetric) model, subscript $l$, the former model is a better representation (with more predictive power) of the input image data than the latter

$$G\text{-}AIC(S_m) < G\text{-}AIC(S_l)$$ (5).

The rational/objective geometric model selection strategy is to minimize the G-AIC values (rather than only the sums of squared residuals) for a whole set of geometric models by means of repeated applications of inequality (5). As there are two models, $S_m$ and $S_l$, in (5), one sets this inequality up for non-disjoint pairs of geometric models, one at a time, and tests if the inequality is fulfilled.

The geometric model-bias correction term 2 $(dN + n)$ $\hat{\varepsilon}^2$ in equation (4) will for a less constrained model be larger than its counterpart for a more constrained model (with equal $N$ and $d$). In other words, the better fitting, less constrained, model features a higher "geometric model selection penalty" than its worse fitting, more constrained, counterpart. This kind of interplay between fitting the input image data better at the expense of a higher model selection penalty provides the basis for objective geometric model selections by minimizing their G-AIC values over a complete set of geometric models.

The fulfillment of inequality (5) allows for a more constrained/symmetric model of the input image data to be selected in a statistically sound manner as a better representation of the input data although its numerical fit, as measured by its sum of squared residuals, is worse than that of the less constrained/symmetric model. Note that the identification of which of the two geometric models is the better representation of the input image data is based solely on the input data itself and the underlying mathematics of Kanatani's theory.

There is no arbitrarily set threshold for the identification of the better model in the presence of a symmetry inclusion relationship, just an inequality that needs to be fulfilled numerically. All of the other crystallographic symmetry classification methods that were so far used in electron crystallography (Hovmöller, 1992, Valpuesta and co-workers, 1994, Wan and co-workers, 2003, Kilaas and co-workers, 2005, Gipson and co-workers, 2007, Zou and co-workers, 2011) and the computational symmetry community (Liu and co-workers, 2009) feature such thresholds.

At first sight, it would seem that estimates of $\hat{\varepsilon}^2$ are needed to make objective geometric model selections by the minimization of their G-AIC values by means of inequality (5) and the definition of the first-order model selection criterion (4). Each geometric model features a different presumed geometric information content, on the one hand, and presumed non-information (generalized noise) content, on the other hand.

There are, however, workarounds to estimating $\hat{\varepsilon}^2$ that not only identify the best possible separation of geometric information and non-information, but also give an estimate of the prevailing noise in the input image data. The two workarounds take in this paper advantage of both the translationengleiche symmetry inclusion relationships between plane symmetry groups as shown in Figure 4a and the symmetry inclusion relationships between the 2D point groups that are projected Laue classes as shown in Figure 4b, i.e. non-disjointness in other words.

For crystallographic symmetry classifications of more or less 2D periodic images, the dimension of the geometric models is zero (as the data are the intensities of individual pixels, which are considered to be zero-dimensional, i.e. points). The degrees of freedom of the geometric models in this paper depend on the number of non-translational symmetry operations in the plane symmetry groups to which the translation-averaged input image data has been symmetrized. They are obtained by the ratio

$$n = \frac{N}{k}$$ (6),

where $k$ is the number of non-translational symmetry operations, which is equal to the multiplicity of the general position per lattice point in all plane symmetry groups. (This number is also one of the two ordering principles of Figures 4a and 4b.)

Equation (6) and what follows from it are good approximations when $N$ is large[A4] (as in this paper). A necessary but not sufficient precondition for $N$ being large in Fourier space is a digital representation of the image to be classified by a very large number of individual pixels in direct space. A complex translation periodic motif with sharp edges and strong contrast changes will produce a large number of complex Fourier coefficients when Fourier transformed.

As already mentioned above, the number of non-translational symmetry operations, $k$ in (6), is one of the two ordering principles of the hierarchy tree of the translationengleiche plane symmetry groups, Figure 4a. This number is given both on the left and right hand side of this figure and increases from the bottom to the top of the symmetry hierarchy tree. The other ordering principle in this figure is the non-disjointness of maximal subgroups and minimal supergroups of the plane symmetry groups specified for their crystallographic settings. These symmetry inclusion relations are in Figure 4a marked by arrows between maximal subgroups and minimal supergroups that are translationengleich. The ratios of the sums of squared residuals of the complex structure-bearing Fourier coefficients for "climbing up" from a lower level (subscript $l$ for *less-symmetric*) of the hierarchy to a higher level (subscript $m$ for *more-symmetric*) that is permitted by the fulfillment of inequality (5) for the special case of equal numbers of complex Fourier coefficients of the lower and higher symmetric geometric model of the input image data ($N_m = N_l$) are also given in Figure 4a.

Translationengleich in the previous paragraph means that the addition of a non-translational symmetry operation to the unit cell of a lower symmetric group, which has the status of a maximal subgroup, results in a unit cell of a higher symmetric group, which is the former's minimal supergroup. Changes from a primitive unit cell to a centered unit cell and vice versa are permitted (Burzlaff and co-workers, 1968), as they represent, effectively, orientation changes of symmetry operations with respect to the conventional unit cell vectors. Analogous considerations apply to the hierarchy of the projected 2D Laue classes, where there are per definition only point symmetries to consider.

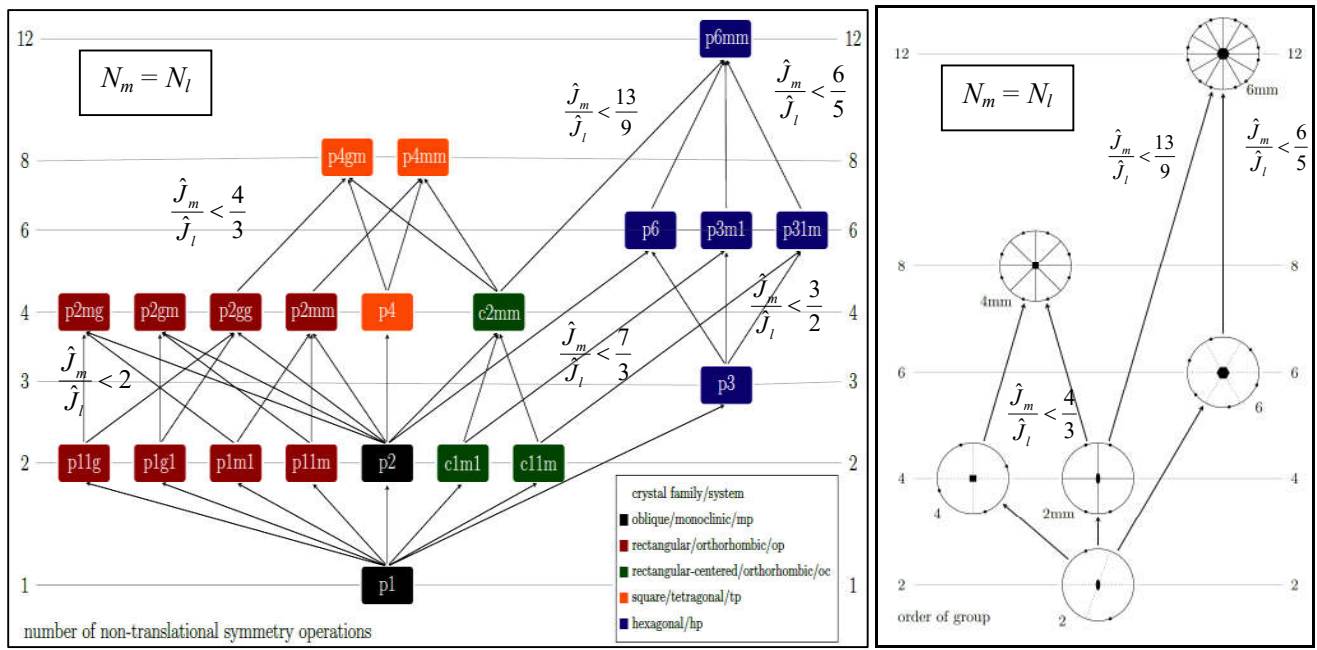

**Figure 4. (a-left)** Hierarchy tree of the translationengleiche plane symmetry groups with ratios of sums of squared complex Fourier coefficient residuals as insets. **(b-right)** Hierarchy tree of the crystallographic 2D point groups that are projected Laue classes. The inset ratios of the sums of squared residuals are valid for equal numbers of structure-bearing Fourier coefficients of geometric models and apply to transitions from a certain $k_l$ level of the graph to a permitted $k_m$ level. Subscript $l$ in these ratios stands for *less-symmetric/constrained* and subscript $m$ stands for *more-symmetric/constrained*. Maximal subgroups are connected to their minimal supergroups by arrows in both parts of this figure.

The translation averaged geometric model of some input image data (with plane symmetry group *p1*) is, for example, non-disjoint from the *c1m1* symmetrized model of these data, as that plane symmetry group is a minimal supergroup of *p1*. The centered plane symmetry group *c1m1* with $k = 2$ is in turn in a maximal subgroup relationship with plane symmetry group *p3m1* with $k = 6$, see Figure 4a. Whenever there is no connecting arrow between two plane symmetry groups in Figure 4a and two projected Laue classes in Figure 4b, that pair of symmetry groups is disjoint.

The two ordering principles in Figure 4b are analogous to those in Figure 4a. The order of the 2D point group/projected Laue class on the left and right hand side of the hierarchy tree increases from the bottom to the top of the hierarchy tree.

Maximal subgroups are connected to their minimal supergroups by arrows. The ratios of the sums of squared residuals of the amplitudes of the structure-bearing Fourier coefficients for climbing up from a lower level of the hierarchy to a permitted higher level of the 2D point groups are also given in this figure for $N_m = N_l$. For an analogous pair of geometric models with hierarchy levels $k_m$ and $k_l$, the same ratios of squared residuals are given in both parts of Figure 4. This is because the same inequalities are applicable for climbing up tests in both hierarchy trees.

In the above-mentioned workarounds to estimating $\widehat{\varepsilon}^2$, one sets up inequality (5) for two non-disjoint models of the input image data that were symmetrized to non-disjoint plane symmetry groups,

and takes advantage of the estimate

$$\widehat{\varepsilon}_l \approx \sqrt{\frac{\widehat{J}_l}{r_l \, N - n_l}} \qquad (7a),$$

for the amount/level of approximately Gaussian distributed noise in the lower symmetric model (designated by the subscript $l$). The variable $r_l$ stands in this estimate for the so-called co-dimension in Kanatani's framework. (In our case, the co-dimension is equal to unity[A5], just as $r_{best}$ in equation (7b) below.)

As long as inequality (5) is fulfilled, one is allowed to climb up in the hierarchy trees of Figure 4. One always starts with the lower symmetric model that corresponds to the anchoring group or class.
Inequality (5) is fulfilled under the conditions:

$$\frac{\widehat{J}_m}{\widehat{J}_l} < 1 + \frac{2(d_l - d_m)N + 2(n_l - n_m)}{r_l N - n_l} \qquad (8a)$$

and

$$\frac{\widehat{\varepsilon}_m^2}{\widehat{\varepsilon}_l^2} < \frac{(2(d_l - d_m) + r_l)N + n_l - 2n_m}{r_m N - n_m} \qquad (8b).$$

So far, we followed Kanatani's general derivation in the *"Model comparison by AIC"* section of his monograph (2005) closely. Now we turn to our specific case of crystallographic symmetry classifications of more or less 2D periodic patterns. For our case[A5], with $d_m = d_l = 0$, $r_m = r_l = 1$, and (6), we obtain from (8a)

$$\frac{\widehat{J}_m}{\widehat{J}_l} < 1 + \frac{2(k_m - k_l)}{k_m(k_l - 1)} \qquad (9a),$$

when the number of data points in both the more and the less symmetric geometric model is the same, $N_m = N_l$. This problem specific inequality is a special case of the general inequality (5) for rational/objective geometric model selections.

For the purpose of this paper, we need a generalization of (9a) for the $N_m \neq N_l$ case of the geometric models that we want to compare with respect to their predictive power. This is because we want to compare our crystallographic symmetry classification results directly to the suggestions that the CRISP program provides, working with the same numerical representations of the geometric models for the input image data that this program allows one to export. Such a generalization of inequality (9a) is provided in Dempsey and Moeck (2020):

$$\frac{\widehat{J}_m}{\widehat{J}_l} < 1 + \frac{2(k_m - \frac{N_m}{N_l} k_l)}{k_m(k_l - 1)} \qquad (9b),$$

and it will be used throughout the rest of this paper.

Note that per inequality (9b), climbing up from the translation-averaged-only model of the input image data to all geometric models that have been symmetrized to minimal supergroups of $p1$ is impossible, as $k_l = 1$ in all of these cases. (There is also a zero sum of squared complex Fourier coefficient residuals for the translation-averaged-only model, equation (1), so that there is no inconsistency.)

One, therefore, simply assumes that there is more than translation symmetry in the input image data and uses inequality (9b) with $k_l = 2$ and 3 as minimum. After having made that assumption, one proceeds with determining what symmetry operations there are in the input image data and to what plane symmetry group they combine.

One needs to carefully distinguish between genuine plane symmetry groups and possibly existing Fedorov type pseudosymmetry groups in the input image data based on the model pair's $\widehat{J}_m$, $\widehat{J}_l$, $k_m$, and $k_l$ values and $N_m$ to $N_l$ ratio. Based on the definitions in the section 1.2 of this paper, the least broken symmetry at the $k_l = 2$ and 3 levels is the first genuine symmetry that is identified and all other genuine symmetries need necessarily be anchored to this particular symmetry group.

In practice, one begins an objective plane symmetry classification by calculating the sums of squared residuals for all of the geometric models that feature a multiplicity of the general position per lattice point (number of non-translational symmetry operations) of two and three, see Figure 4a. (Note that plane symmetry groups $c1m1$ and $c11m$ feature two non-translational symmetry operations each, the multiplicity of the general position in the centered unit cell is four, but there are two lattice points per unit cell.)

All of the geometric models with two and three non-translational plane symmetry operations are disjoint from each other per definition. Combinations of the groups with two and three non-translational plane symmetry operations lead to the majority of plane symmetry groups that are higher up in the hierarchy tree, Figure 4a.

When there is more than translation symmetry in the input image data, at least one of the geometric models that have been symmetrized to a plane symmetry group with two or three non-translational symmetry operations will have a low sum of squared residuals of the complex structure-bearing Fourier coefficients. The plane symmetry group of that model is necessarily non-disjoint from its minimal supergroups so that tests of whether a climbing up in the plane symmetry hierarchy tree is allowed by inequality (9b) can proceed until the Kullback-Leibler best geometric model of the image input data has been found.

By first calculating the sums of squared residuals for all eight geometric models of the input image data that feature $k = 2$ and 3, we make sure we know from which plane symmetry group the anchoring and climbing up in the hierarchy tree of plane symmetry groups, Figure 4a, shall proceed in this paper, as long as permitted by the fulfillment of inequality (9b).

The sums of squared residuals of the complex structure-bearing Fourier coefficients of the geometric models of the input image data that have been symmetrized to higher symmetric plane symmetry groups may be calculated on an as-needed basis. Note that the whole procedure can be programmed and does not require visual inspections and comparisons of differently symmetrized versions of the input image data. This makes the information theory based classification techniques very different to the other plane symmetry classification methods that are used in contemporary electron crystallography.

Note that to conclude that a certain minimal supergroup is a plane symmetry that minimizes the G-AIC value of a geometric model of the image input data within a set of models, inequality (9b) has to be fulfilled for all maximal subgroups (and in turn their maximal subgroups). If that is not the case, that plane symmetry is only a Fedorov type pseudosymmetry as it is broken to a larger extent than the genuine plane symmetry that the hypothetical noise-free version of the input image most likely possesses. The formally correct crystallographic symmetry classification of a more or less 2D periodic pattern is the plane symmetry group and projected Laue class that minimize the respective G-AIC values.

In the case of projected Laue classes, there is a zero sum of squared structure-bearing Fourier coefficient amplitude residuals for point symmetry group $2$, see equation (2), because the Fourier transform is centrosymmetric. The anchoring group is, therefore, to be found at the $k_l = 4$ or 6 levels of the hierarchy tree in Figure 4b. All other considerations for finding the K-L best projected Laue class are analogous to those for finding the K-L best plane symmetry group.

For consistent crystallographic symmetry classifications of more or less 2D periodic patterns, the K-L best projected Laue class and the K-L best plane symmetry group need to be compatible with each other as they are based on complementing aspects of the same input image data. As the example of the noisiest classified crystal pattern below will show, it is possible that the formally correct K-L best plane symmetry group and K-L best projected Laue class are crystallographically incompatible with each other. When this happens, it signifies a partial breakdown of the information theoretic methodology that results from equation (4) being no longer a good approximation of equation (3) and/or the generalized noise not being Gaussian distributed to a sufficient approximation.

A good estimate of the variance of the generalized noise, which needs to be approximately Gaussian distributed, can be obtained *after* the correct crystallographic symmetry classification has been made, i.e. the K-L best model in the set has been identified, from

$$\widehat{\varepsilon}_{best}^2 \approx \frac{\widehat{J}_{best}}{r_{best} N_{best} - n_{best}} \tag{7b},$$

where the subscript *best* stands for the Kullback-Leibler best model of the input image data. This estimate is in the same format as (7a), i.e. the representation of the estimated square of the noise level of the geometric model that features the lower symmetric group or class in a pair-wise model comparison procedure. When the K-L best model of the input image data has been identified, there is obviously no further climbing up allowed in the symmetry hierarchy trees of Figure 4. This is because the G-AIC values inequality (5) can no longer be fulfilled using inequalities (8a) and (8b) as well as (9a) or (9b).

The estimate in (7b) is needed for calculations of geometric Akaike weights of a set of geometric models for the input image data. These weights are the probabilities that a certain geometric model of the input image data is indeed the K-L best model in a set of geometric models. They are to be calculated on the basis of the G-AIC values according to equation (4) with (7b) for the noise term. This is not done in this paper and the reader is referred to Moeck (2018) and Dempsey and Moeck (2020) for details on how likelihoods of geometric models are transformed into model probabilities. Providing geometric Akaike weights is a route to deriving uncertainty measures for plane symmetry group and projected Laue class classifications, without which crystallographic symmetry measurements, i.e. quantifications, are simply incomplete (Helliwell, 2021). Another route to deriving classification uncertainty measures is to use $N_m \neq N_l$ generalizations of the confidence level equations for selecting minimal supergroups over their maximal subgroups, see appendix B.

Note that to obtain reasonable results for the geometric Akaike weights, a normalization of the residuals, as described in Dempsey and Moeck, 2020, is mandatory when one works with *.hka files from the CRISP program. We use the same normalization in this paper as it is inconsequential for the ranking of geometric models by their G-AIC values.

## 4. Objective crystallographic symmetry classifications of three synthetic crystal patterns and an optimal crystallographic image processing induced noise suppression

### 4.1. Details of the classification procedure as employed in this paper

As already mentioned in the introductory section 1.7 to this paper, crystallographic symmetry classifications are done here with both the author's methods and the electron crystallography program CRISP (Hovmöller, 1992, Zou and co-workers, 2011, Zou and Hovmöller, 2012) using the same *.hka files[A6] of the latter program. An appropriately chosen series of these files contain all of the information on the structure-bearing Fourier coefficients of the differently symmetrized geometric models of the input image data that is needed for objective classification into plane symmetry groups and projected Laue classes.

In the CRISP program, these files are internally used to calculate symmetry deviation quantifiers in the form of sets of normalized amplitude and phase angle differences of symmetrized structure-bearing complex Fourier coefficient sets of the input image data with respect to the structure-bearing complex Fourier coefficient set of this data itself. (Ratios of sums of odd to even Fourier coefficient amplitudes are also calculated from these files when they are meaningful.) The *.hka files are also used internally to create symmetrized direct space versions of the input image data by Fourier back transforming for visual comparisons by the CRISP program's user.

These files can be interactively edited in CRISP. This allows, for example, for restrictions of the geometric models of the input image to a desired dynamic range of the Fourier coefficient amplitudes. The program's default value for this dynamic range is 200. (The maximal amplitude is always set to 10,000.)

Lowering the dynamic range leads to a reduction of the number of complex structure-bearing Fourier coefficients of the geometric models, and we will make use of that for both the noise-free and the modest amount of added noise pattern in the analyzed series of crystal patterns, see Figures 1 and 5.

Calculating the discrete Fourier transform with CRISP in its maximal dynamic range setting resulted in 3,666 complex structure-bearing Fourier coefficients for the translation averaged model of the undisturbed crystal pattern in Figure 1. The patterns in Figures 2 and 3 are, on the other hand, restricted to the back-transform of the strongest 956 complex Fourier coefficients without any symmetrizing.

A limited dynamic range of the Fourier coefficient amplitudes may lead to a reduction in the accuracy of the geometric models of the input image data. As the direct visual comparison of the crystal patterns in Figures 1 and 2 suggests, this is not a problem in the present study. Limiting the dynamic range has, on the other hand, the benefit of reducing "Fourier ripples" around features with very strong contrast changes, as can be seen in Figure 2.

With a very large number of data points in the discrete Fourier transform of some input image data with very small amplitudes, one has to wonder if the accuracies of geometric models of the input image data are not compromised by the limited representation length of real numbers in a computer program, accumulated rounding errors, and numerical approximations in the calculation of the discrete Fourier transform.

The CRISP program also allows for restrictions of the spatial resolution of the geometric models of the input image data in reciprocal space. This spatial resolution is akin to the Abbe[1] resolution. Restricting the spatial resolution is typically necessary for noisy crystal patterns that are to be classified and will be done here as well for both of the noisy images, Figures 5 and 6. What will be called "spread noise" below is particularly effective in reducing the number of well-resolved data points in a discrete Fourier transform, as demonstrated by Dempsey and Moeck (2020). Without judicious restrictions of the dynamical range of the structure-bearing Fourier coefficient amplitudes and the Abbe resolution of a noisy crystal pattern, one may produce conspicuous artifacts in the subsequent crystallographic processing of the more or less 2D periodic image.

The MATLAB script hkaAICnorm, as written by a graduate student of this author (Dempsey and Moeck, 2020), was used for the extraction of the pertinent information from the exported *.hka files. That script can be freely downloaded (Program III) and calculates the sums of normalized squared residuals for all of the geometric models that are used in this study from a series of *.hka files from the CRISP program. (As described in Dempsey and Moeck (2020), the script works with normalized amplitudes of the structure-bearing Fourier coefficients in order to keep the numbers in the data tables below small.)

The noise-free pattern, Figure 1, of the synthetic crystal pattern series is classified with respect to its plane symmetry group and projected Laue class in the following subsection. The third subsection presents the classifications of the two noisy patterns, Figures 5 and 6, of the series.

The results of the crystallographic processing of the two noisy patterns of the crystal pattern series are given in the fourth subsection.

## 4.2. Classification of the noise-free pattern in the series of crystal patterns

Table 1 lists the sums of squared residuals for a judicious selection of geometric models of the noise-free pattern, of which a small section is shown in Figure 1. In all three analyses of this paper, circular area selections with a diameter of 1024 pixels were made in direct space for the calculation of the discrete Fourier transforms. These sections contained approximately 88 primitive unit cells of the crystal patterns that are to be classified.

No explicit spatial restriction was made in Fourier space for the calculation of the entries in Table 1 as it is considered to be free of generalized noise. The dynamic range of the Fourier coefficient amplitudes was set to 100 in order to restrict the number of data points $N$ in inequality (9b) to something that is easier managed. (This amounts to an implicit spatial resolution restriction.)

Note that the first seven entries in this table consist of the geometric models of the input data that feature two non-translational symmetry operations, whereas the $8^{th}$ entry features three such operations. All of these eight models are disjoint from each other (and there are no connecting vectors between them in the plane symmetry hierarchy tree in Figure 4a).

The subsequent three entries in Table 1 consist of geometric models that feature four non-translational symmetry operations. The last two entries feature eight such operations and the two corresponding models are disjoint from each other (in the translationengleiche sense, Burzlaff and co-workers, 1968).

The lowest sum of squared residuals of the complex Fourier coefficients is for the crystal pattern that underlies Figure 1 obtained for the geometric model that has been symmetrized to plane symmetry group *p2*, see Table 1. The geometric model with plane symmetry group *p4* is listed in this table as the one that has the lowest (non-zero) sum of squared residuals of the amplitudes of the Fourier coefficients.

**Table 1:** Results of the hkaAICnorm MATLAB script on the noise-free pattern that underlies Figure 1 for geometric model selections by G-AIC value minimization using inequality (9b).

| Plane symmetry group to which the image data have been symmetrized | Sum of squared residuals of complex Fourier coefficients | Sum of squared residuals of Fourier coefficient amplitudes | Number of Fourier coefficients in the geometric model of the image data |
|---|---|---|---|
| *p2* | 0.0042 | None | 956 |
| *p1m1* | 1.8799 | 0.0052 | 937 |
| *p11m* | 1.8642 | 0.0052 | 937 |
| *p1g1* | 0.0094 | 0.0052 | 934 |
| *p11g* | 0.0081 | 0.0052 | 934 |
| *c1m1* | 0.0103 | 0.0053 | 924 |
| *c11m* | 0.0110 | 0.0053 | 924 |
| *p3* | 2.5290 | 1.3339 | 954 |
| *p2gg* | 0.0096 | 0.0052 | 931 |
| *c2mm* | 0.0119 | 0.0053 | 924 |
| *p4* | 0.0065 | 0.0021 | 948 |
| *p4mm* | 1.9558 | 0.0063 | 918 |
| *p4gm* | 0.0102 | 0.0061 | 912 |

The symmetry in the amplitude map of the discrete Fourier transform is for the *p4* symmetry model of the input image data point group *4* (Aroyo, 2016, Hahn, 2010), which is a projected Laue class. For easy reference, the entries for geometric models with plane symmetry groups *p2* and *p4* are marked in Table 1 by the shading of the respective two rows.

The selection of entries in Table 1 has been made in order to demonstrate the climbing up from a lower level of the hierarchy of plane symmetry groups, see Figure 4a, to the next higher level. The tests if such a climbing up is allowed by the fulfillment of inequality (9b) always start at the geometric model with the plane symmetry that has the lowest sum of squared residuals of the complex Fourier coefficients amongst the mutually disjoint models with two and three non-translational symmetry operations, i.e. the anchoring group. That starting model features always per definition a genuine symmetry, but more genuine symmetries can potentially be identified by the fulfillment of inequality (9b) for some of its non-disjoint models that may combine with the first identified genuine symmetry to some higher-level genuine symmetry.

As already mentioned above, the geometric model that was symmetrized to plane symmetry group $p2$ features the lowest squared residual of the complex Fourier coefficients in Table 1. Symmetry models that are candidates for climbing up from the geometric model that was symmetrized to $p2$ in the plane symmetry group hierarchy tree, Figure 4a, e.g. $p2mg$, $p2gm$, $p2gg$, $p2mm$, $c2mm$, or $p4$ need to have a sufficiently small sum of squared residuals (and G-AIC values) with respect to all of their maximal subgroups in order to be declared genuine. Otherwise, they are Fedorov type pseudosymmetries by definition. Geometric models of the input image data with low (but not the lowest) sums of squared complex Fourier coefficient residuals and two or three non-translational symmetry operations may either reveal a genuine symmetry or a Fedorov type pseudosymmetry.

Plane symmetry group $p4$ has only one maximal subgroup, i.e. $p2$, so that only one inequality fulfillment test is needed to find out if the former is a genuine symmetry of the crystal pattern in Figure 1 or not. For each of the other five geometric models mentioned in the previous paragraph, one would need to complete three inequality fulfillment tests. It is, however, already quite clear from the entries in Table 1 that only the models that were symmetrized to plane symmetry groups $p1g1$, $p11g$, $c1m1$, and $c11m$, have reasonably low sums of squared residuals (and G-AIC values) to make them reasonable candidates for climbing up to geometric models that feature a genuine supergroup that they share with $p2$. The models with plane symmetry groups $p1m1$ and $p11m$ feature very high sums of squared residuals of the complex Fourier coefficients in Table 1 so that it is unreasonable to expect they could possibly combine with the geometric model that features the $p2$ anchoring group. The crystal pattern in Figure 1 can, therefore, not be classified as belonging to plane symmetry groups $p2mm$, $p2gm$, and $p2mg$. Analogously, given that the entry in the second column of Table 1 is even higher for the geometric model that was symmetrized to plane symmetry group $p3$, the pattern in this figure is definitively not hexagonal.

Table 2 gives the ratios of the sums of squared residuals of the complex Fourier coefficients for the non-disjoint models of Table 1 (left-hand side of inequality (9b) in the second column) together with the maximal value that these ratios may have (right-hand side of inequality (9b) in the third column) in the context of minimization of the G-AIC value of the higher symmetric model of a pair of non-disjoint geometric models of the input image data. The tests if climbing up to the next level of the plane symmetry hierarchy tree is allowed consist in a simple comparison of the numerical values in the second and third column of Table 2, which is recorded in the fourth column.

There is only one unconditional "yes" in the fourth column of this table, as marked by the shading of the corresponding row, so that the conclusion has to be drawn that the geometric model which has been symmetrized to plane symmetry group $p4$ features the only other genuine symmetry in the crystal pattern that underlies Figure 1, i.e. the noise-free pattern of the series.

**Table 2:** Numerical values of ratios of sums of squared residuals of the complex Fourier coefficients of non-disjoint models of the noise-free pattern, Figure 1, that are either within their maximal allowance or not.

| | Left-hand side of (9b) | Right-hand side of (9b) | Inequality (9b) fulfilled? |
|---|---|---|---|
| $p2gg$ over $p2$ | 2.285714 | 2.0261506 | no, blocking ascent |
| $p2gg$ over $p1g1$ | 1.021277 | 2.0032312 | yes, but due to pseudosymmetry |
| $p2gg$ over $p11g$ | 1.185185 | 2.0032312 | yes, but due to pseudosymmetry |
| $c2mm$ over $p2$ | 2.83333 | 2.0 | no, blocking ascent |
| $c2mm$ over $c1m1$ | 1.155340 | 2.0 | yes, but due to pseudosymmetry |
| $c2mm$ over $c11m$ | 1.081818 | 2.0 | yes, but due to pseudosymmetry |
| $p4$ over $p2$ | 1.547619 | 2.008368 | yes |
| $p4mm$ over $p4$ | 300.8923 | 1.3438819 | no, blocking ascent |
| $p4gm$ over $p4$ | 1.569231 | 1.3459916 | no, blocking ascent |
| $p4gm$ over $p2gg$ | 1.06250 | 1.3401361 | yes, but due to pseudosymmetry |
| $p4gm$ over $c2mm$ | 0.857143 | 1.3376623 | yes, but due to pseudosymmetry |

It is important to realize that all genuine symmetries above the $k = 2$ and 3 level must by definition be anchored to the least broken plane symmetry group, i.e. the one with the lowest sum of squared residuals for the complex Fourier coefficients at the $k_l = 2$ and 3 levels in Figure 4a. The fulfillment of inequality (9b) for a pair of non-disjoint geometric models that does not fulfill this overwriting requirement can per definition only signify a Fedorov type pseudosymmetry.

The "strength" of a Fedorov type pseudosymmetry correlates inversely with the sum of the squared residuals of the complex Fourier coefficients of its corresponding geometric model of the input image data. Plane symmetry groups $p2gg$ and $c2mm$ must be Fedorov type pseudosymmetries of the crystal pattern in Figure 1 because climbing up from $p2$ is not permitted, see first and fourth entry in Table 2. These two plane symmetry groups are strong Fedorov type pseudosymmetries because the sums of squared complex Fourier coefficients residuals of the corresponding two geometric models of the input image data are low in Table 1. Their maximal subgroups $p1g1$, $p11g$, $c1m1$, and $c11m$ are even stronger Fedorov type pseudosymmetries as they are disjoint from the $p2$ anchoring group and the corresponding geometric models feature lower sums of squared residuals of the complex Fourier coefficients in Table 1 than the models that represent the minimal supergroups $p2gg$ and $c2mm$.

Note that climbing-up tests for strong Fedorov type pseudosymmetries to the $k_m = 4$ level, i.e. $p2gg$ and $c2mm$, and up to $k_m = 8$, i.e. $p4gm$, result in rather low values for the left-hand side of inequality (9b) in Table 2. This is due to the corresponding sums of squared complex Fourier coefficient residuals for the matching $k_l = 2$ and 4 levels being on the same order in Table 1. The ratios of such sums may, for strong Fedorov pseudosymmetries, even fall below unity[A7], as shown for the last entry in Table 2.

The identification of the projected Laue class that minimizes the G-AIC value for the crystal pattern that underlies Figure 1 proceeds analogously. Laue class $4$ has already been identified above as the point symmetry of the amplitude map of the geometric model that has been symmetrized to plane symmetry group $p4$. Because the $p4$ model has the lowest squared Fourier coefficient amplitude residual sum in Table 1, point group $4$ is the anchoring point group for the projected Laue class classification of the crystal pattern in Figure 1. Both this projected Laue class and 2D Laue class $2mm$ feature four point symmetry operations, $k_l = 4$, and are disjoint from each other, see point group hierarchy tree in Figure 4b.

Table 3 gives the ratios of the sums of the squared Fourier coefficient amplitude residuals for the non-disjoint models of Table 1 (with $k_l = 4$) together with the maximal value that these ratios may have for a climbing up to the $k_m = 8$ level. Obviously, one cannot climb up from the model with projected Laue class $4$ to the non-disjoint model with projected Laue class $4mm$ with $k_m = 8$ (in Figure 4b), based on the numbers in this table.

Based on the low sums of squared Fourier coefficient amplitude residuals in Table 1, the models for projected Laue classes $2mm$ and $4mm$ reveal pseudosymmetries in the input image data. This is fully consistent with the identified Fedorov type pseudosymmetries at the plane symmetry group level.

**Table 3:** Numerical values for the ratio of the sums of squared Fourier coefficient amplitude residuals of non-disjoint geometric models of the noise free pattern, Figure 1, that are either within their maximal allowance or not.

|  | Left-hand side of inequality (9b) | Right-hand side of inequality (9b) | Inequality fulfilled? |
|---|---|---|---|
| *4mm* over *4* (in c2mm setting) | 3 | 1.3438819 | no, as it should |
| *4mm* over *4* (in p2gg setting) | 2.90476 | 1.3577236 | no, as it should |
| *4mm* over *2mm* (in p2gg setting) | 1.2115385 | 1.3379878 | yes, but due to pseudosymmetry |
| *4mm* over *2mm* (in c2mm setting) | 1.1886792 | 1.3246592 | yes, but due to pseudosymmetry |

To conclude this subsection: plane symmetry group $p4$ (which contains $p2$ as its only maximal subgroup) and projected Laue class $4$ are identified as both genuine in the crystal pattern that underlies Figure 1 and crystallographically consistent with each other. The identified Fedorov type pseudosymmetries at the lowest level of the hierarchy tree of plane symmetry groups are $p1g1$, $p11g$, $c1m1$, and $c11m$. These pseudosymmetries combine with each other and the identified genuine symmetries to the pseudosymmetry groups $p2gg$, $c2mm$, and $p4gm$. There are corresponding $2mm$ and $4mm$ pseudosymmetries in the Fourier transform amplitude map of the noise-free crystal pattern in Figure 1, but no $4mm$ pseudo-site symmetry in the direct space unit cell of the input image data since the $p1m1$ and $p11m$ models of this data feature sums of squared complex Fourier coefficient residuals that are way too large to pass climbing up tests in the plane symmetry hierarchy tree of Figure 4a.

### 4.3. Classifications of the two noisy patterns of the series of crystal patterns

Figures 5 and 6 show sections of the two synthetic patterns that were obtained by adding approximately Gaussian distributed noise to the crystal pattern that served as the basis of Figure 1, i.e. the approximately 144 periodic motif repeats containing expanded representation of the original graphic artwork (Knoll, 2003) that is considered to be free of generalized noise. The freeware program GIMP (Program I) was used to add the noise.

Spread noise swaps individual pixel intensities in the horizontal and vertical directions by a selected number of pixels[A8]. Strictly Gaussian distributed noise only changes the individual pixel values but not their positions in the translation periodic unit cell. The employed mixtures of strictly Gaussian distributed noise and spread noise add up to approximately Gaussian distributed noise. (The strictly Gaussian distributed noise had been added to the crystal pattern in Figure 1 before the spread noise was added with GIMP.)

The effects of the added noise are clearly visible in Figures 5 and 6 and their histogram insets when compared to the histogram inset in Figure 1 and that figure itself. Compared to Figure 5, there is approximately five times as much added noise in Figure 6.

We classify the noisy crystal pattern in Figure 5 first. The dynamic range in the employed *.hka files from CRISP was set to 100. The selection in Fourier space was set to a 350 pixel radius (out of the maximal possible 512 pixel radius). The combination of both of these settings resulted in a reasonable number of Fourier coefficients in the last column of Table 4. A consequence of these two settings is a contrast reduction of the crystallographically processed version of this pattern, Figure 7 (in the fourth subsection below), with respect to the crystal pattern in Figure 1. These settings ensured, on the other hand, that there are only very minor processing artifacts in the pattern of Figure 7.

The geometric model with plane symmetry group *p2* features again the lowest sum of squared residuals of the complex Fourier coefficients in this table. Also as before, the model that was symmetrized to plane symmetry group *p4* features the lowest sum of Fourier coefficient amplitude residuals. Again, the rows for these two geometric models of the input image data are highlighted in Table 4 for easy reference by shading.

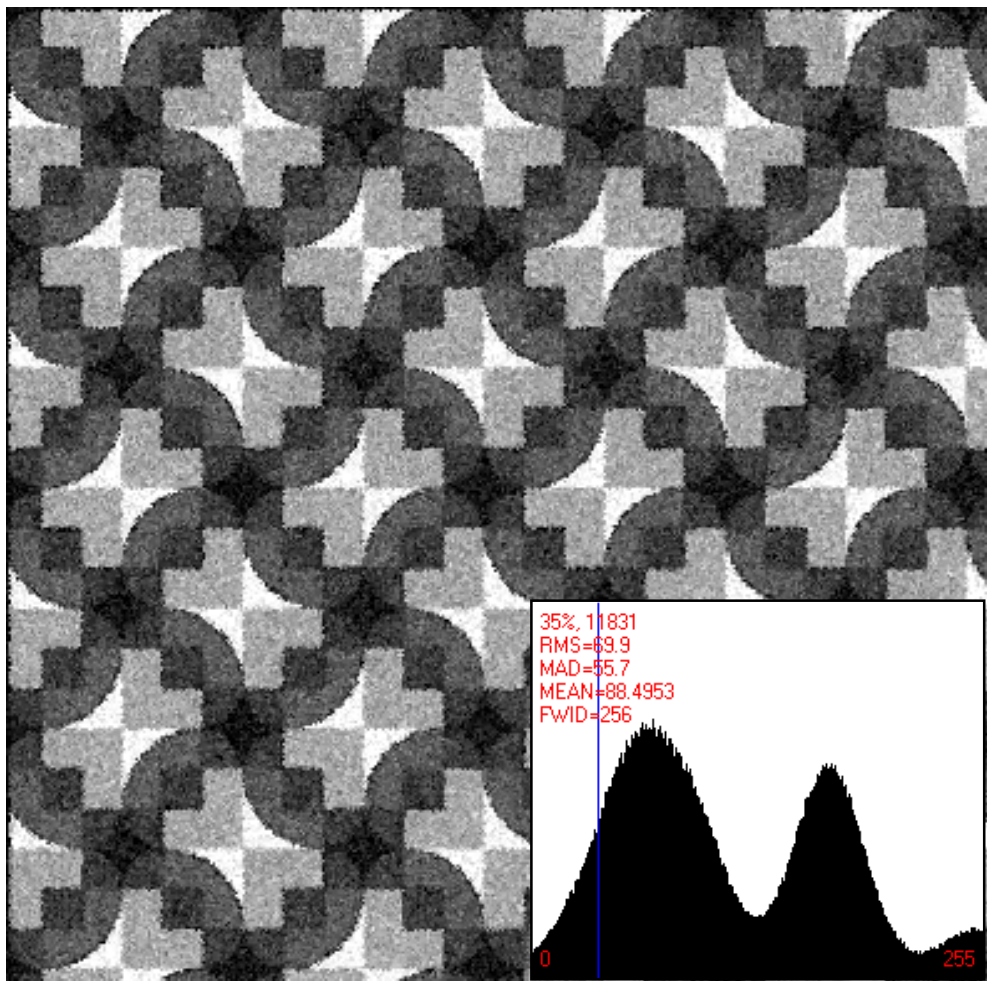

**Figure 5.** Section of the underlying crystal pattern of Figure 1 with a moderate amount of approximately Gaussian distributed noise added. The histogram of the whole pattern is provided as inset. Note that there are only three broad peaks in this histogram, whereas the noise-free histogram of Figure 1 features five narrow peaks.

Analogous to Table 2, Table 5 gives the ratios of the sums of the squared residuals of the complex Fourier coefficients for climbing up tests. There are four unconditional "yes" in Table 5 when the prior information on the objective symmetry classification of the noise-free pattern of the crystal pattern series from the previous subsection is not used. The rows of the corresponding entries are again marked by shading.

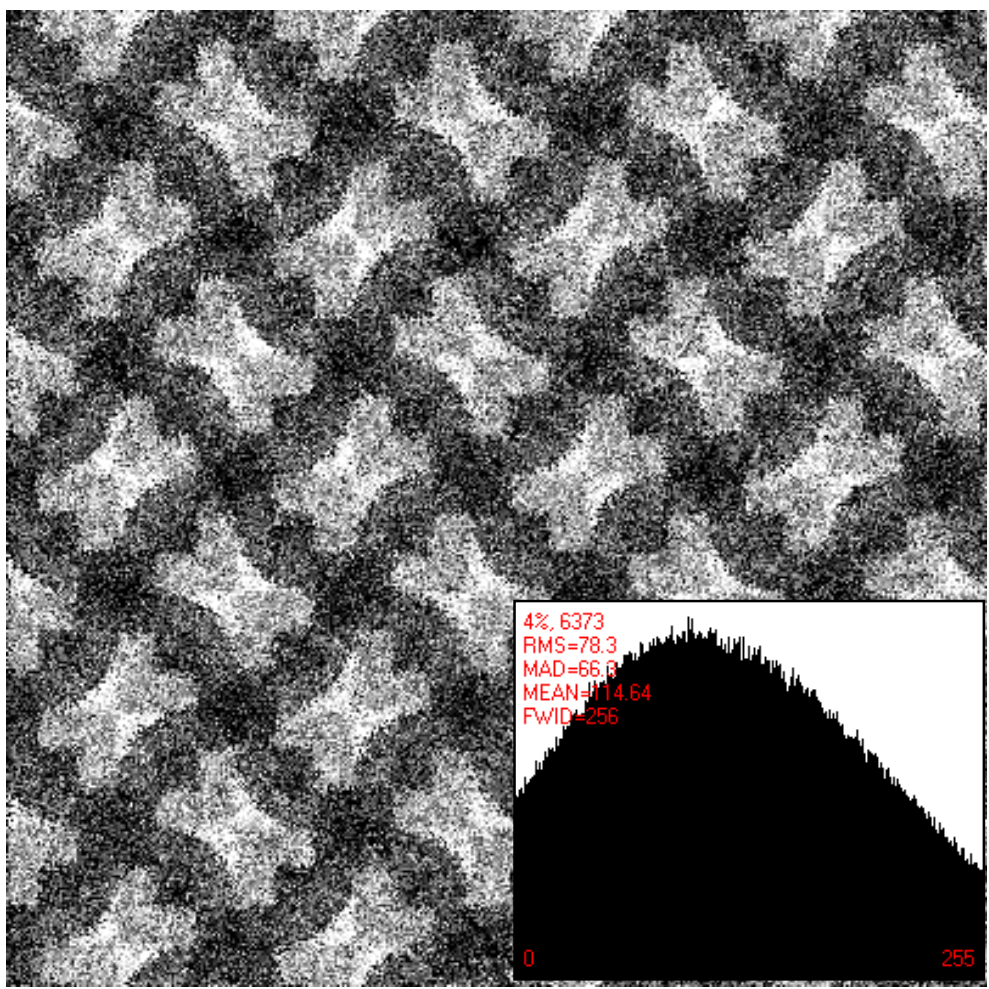

**Figure 6.** Section of the underlying crystal pattern of Figure 1 with a large amount of approximately Gaussian distributed noise added. The histogram of the whole pattern is provided as inset. Note that all of the five narrow peaks in the histogram in Figure 1 are now "overwhelmed" by the added noise, resulting in a single peak that may be characterized as approximately Gaussian distribution but with fat tails[A9].

The preliminary conclusion from the shaded rows in Table 5 is that the genuine plane symmetry group of the noisy crystal pattern in Figure 5 must either be *p2gg* or *p4*. Both plane symmetry groups are disjoint from each other, see Figure 4a, so that one of these two groups has to be a Fedorov type pseudosymmetry per definition. The decision about which of these two plane symmetries is genuine relies on the necessity of the crystallographic consistency of the plane symmetry classification with the Laue class classification of the noisy pattern in Figure 5.

**Table 4:** Results of the hkaAICnorm MATLAB script on the modest amount of noise added pattern that underlies Figure 5 for geometric model selection by G-AIC value minimization using inequality (9b).

| Plane symmetry group to which the image data have been symmetrized | Sum of squared residuals of complex Fourier coefficients | Sum of squared residuals of Fourier coefficient amplitudes | Number of Fourier coefficients in the geometric model of the image data |
|---|---|---|---|
| *p2* | 0.0041 | none | 665 |
| *p1m1* | 1.7207 | 0.0041 | 654 |
| *p11m* | 1.7210 | 0.0041 | 654 |
| *p1g1* | 0.0059 | 0.0041 | 652 |
| *p11g* | 0.0066 | 0.0041 | 652 |
| *c1m1* | 0.0081 | 0.0043 | 655 |
| *c11m* | 0.0081 | 0.0043 | 655 |
| *p3* | 2.0554 | 1.3052 | 685 |
| *p2gg* | 0.0066 | 0.0041 | 650 |
| *c2mm* | 0.0102 | 0.0043 | 655 |
| *p4* | 0.0040 | 0.0015 | 648 |
| *p4mm* | 1.7934 | 0.0050 | 644 |
| *p4gm* | 0.0074 | 0.0050 | 640 |

The anchoring Laue class is point symmetry group *4* because the corresponding *p4* symmetrized model of the noisy pattern in Figure 5 features in Table 4 the lowest sum of squared residuals of the Fourier coefficient amplitudes. The point symmetry in the amplitude maps of the discrete Fourier transforms of the geometric models of the crystal pattern in Figure 5 that were symmetrized to plane symmetry groups *p1m1*, *p11m*, *p1g1*, *p11g*, *c1m1*, *c11m*, *p2gg*, and *c2mm* is point symmetry/Laue class *2mm* (Aroyo, 2016, Hahn, 2010).

**Table 5:** Numerical values of ratios of sums of squared residuals of the complex Fourier coefficients of non-disjoint models of the pattern with a moderate amount of added noise, Figure 5, that are either within their maximal allowance or not.

| | Left-hand side of (9b) | Right-hand side of (9b) | Inequality fulfilled? |
|---|---|---|---|
| *p2gg* over *p2* | 1.6097561 | 2.0225564 | yes |
| *p2gg* over *p1g1* | 1.1186441 | 2.0030675 | yes |
| *p2gg* over *p11g* | 1.0 | 2.0030675 | yes |
| *c2mm* over *p2* | 2.4878049 | 2.0 | no, blocking ascent |
| *c2mm* over *c1m1* | 1.2592593 | 2.0 | yes, but due to pseudosymmetry |
| *c2mm* over *c11m* | 1.2592593 | 2.0 | yes, but  due to pseudosymmetry |
| *p4* over *p2* | 0.9756098 | 2.0255639 | yes |
| *p4mm* over *p4* | 448.35 | 1.3353909 | no, blocking ascent |
| *p4gm* over *p4* | 1.85 | 1.3374486 | no, blocking ascent |
| *p4gm* over *p2gg* | 1.1212121 | 1.3384615 | yes, but due to pseudosymmetry |
| *p4gm* over *c2mm* | 0.7254902 | 1.3409669 | yes, but due to pseudosymmetry |

Table 6 is analogous to Table 3 and lists the ratios of sums of squared Fourier coefficient amplitude residuals for the modest amount of added noise pattern in Figure 5. The conclusion from this table is that projected Laue class *4* is the only genuine class as climbing up from the anchoring class to Laue class/point group *4mm* is not allowed. Crystallographically consistent with this is that ascent from the geometric model that was symmetrized to plane symmetry group *p4* to the *p4gm* symmetrized model of the image input data is not allowed, see Table 5.

Note that point symmetry group *4* captures the symmetry in the amplitude map of the discrete Fourier transform of the noisy crystal pattern that underlies Figure 5 better by more than a factor of 2.7 than point group *2mm*, which is at the same $k_l = 4$ level of the hierarchy tree of Figure 4b. It is, therefore, without any doubt the point symmetry of the Kullback-Leibler best geometric model of the amplitude map of that pattern.

Laue class *2mm* is accordingly to Table 6 a pseudosymmetry at the point symmetry level and the corresponding plane symmetry group *p2gg* can also only be a strong Fedorov type pseudosymmetry. With point group *2mm* identified as pseudosymmetry and point group *4* the genuine symmetry in the amplitude map of the discrete Fourier transform of the noisy pattern in Figure 5, there must also be a *4mm* pseudosymmetry in this map. This is confirmed by the numerical values in Table 6.

Note in passing that the ratio of the sums of squared residuals of the complex Fourier coefficients is for the "*p4* over *p2*" row of Table 5 smaller than unity. This is probably the result of both small accumulated calculation errors in the analysis and slight differences in the accuracy of the representation of the geometric models in the employed[A6] *.hka files from CRISP.

There is, again no *4mm* pseudo-site symmetry in the direct space unit cell of that crystal pattern because ascent from the geometric model that was symmetrized to plane symmetry group *p4* to its counterpart with plane symmetry *p4mm* is blocked in Table 5 by a very wide margin.

**Table 6:** Numerical values for the ratio of the sums of squared Fourier coefficient amplitude residuals of non-disjoint models of the moderate amount of added noise pattern, Figure 5, that are either within their maximal allowance or not.

| | Left-hand side of inequality (9b) | Right-hand side of inequality (9b) | Inequality fulfilled? |
|---|---|---|---|
| *4mm* over *4 (in c2mm setting)* | 3.333333 | 1.3353909 | no, as it should |
| *4mm* over *4 (in p2gg setting)* | 3.333333 | 1.3374486 | no, as it should |
| *4mm* over *2mm (in p2gg setting)* | 1.2195122 | 1.3384615 | yes, but due to pseudosymmetry |
| *4mm* over *2mm (in c2mm setting)* | 1.1627907 | 1.3389313 | yes, but due to pseudosymmetry |

Clear distinctions between genuine symmetries and Fedorov type pseudosymmetries were, thus, again obtained. The added approximately Gaussian distributed noise presented no challenge to the crystal pattern classification task with respect to its crystallographic symmetries when the amount of noise was modest.

The preliminary issue which of the two disjoint plane symmetry groups, *p2gg* or *p4*, is the symmetry of the Kullback-Leibler best model of the noisy pattern in Figure 5 was straightforwardly resolved by recognizing point symmetry *4* as the anchoring Laue class. Note that no prior knowledge of the classification of the noise-free pattern in the series of crystal pattern from the previous subsection was used to reach the final conclusions.

As expected, the effect of adding noise is an obscuring of the differences in the amounts of breakings of the various plane symmetry groups. Adding larger amounts of noise that is to a lesser approximation Gaussian distributed should confirm the general trend that genuine symmetries and pseudosymmetries in crystal patterns get more difficult to distinguish. As we will see below, this is indeed the case.

In analogy to Tables 1 and 4, Table 7 gives the characteristics of the geometric models for the rather noisy crystal pattern that underlies Figure 6. All of the sums of squared residuals except those for *p1m1*, *p11m*, *p3*, and *p4mm* are highlighted in this table by shading of the respective rows. This is because, as Table 8 shows, genuine symmetries at the plane symmetry group level can no longer be distinguished from Fedorov type pseudosymmetries as the result of the large amount of added noise.

**Table 7:** Results of the hkaAICnorm MATLAB script on the pattern with a large amount of added noise that underlies Figure 6 for geometric model selection by G-AIC value minimization using inequality (9b).

| Plane symmetry group to which the image data have been symmetrized | Sum of squared residuals of complex Fourier coefficients | Sum of squared residuals of Fourier coefficient amplitudes | Number of Fourier coefficients in the geometric model of the image data |
|---|---|---|---|
| *p2* | 0.0061 | none | 275 |
| *p1m1* | 1.5353 | 0.0039 | 271 |
| *p11m* | 1.5320 | 0.0039 | 271 |
| *p1g1* | 0.0069 | 0.0039 | 265 |
| *p11g* | 0.0078 | 0.0039 | 270 |
| *c1m1* | 0.0085 | 0.0041 | 269 |
| *c11m* | 0.0074 | 0.0041 | 269 |
| *p3* | 1.7565 | 1.2029 | 306 |
| *p2gg* | 0.0098 | 0.0039 | 264 |
| *c2mm* | 0.0115 | 0.0041 | 269 |
| *p4* | 0.0088 | 0.0028 | 276 |
| *p4mm* | 1.5876 | 0.0053 | 276 |
| *p4gm* | 0.0109 | 0.0051 | 266 |

Plane symmetry group *p4gm* is now identified as genuine and the symmetry that most likely underlies the rather noisy crystal pattern that underlies Figure 6. Note that ascent in the plane symmetry hierarchy tree of Figure 4a is now permitted all the way up to the top of the *p4gm* branch since inequality (9b) is fulfilled for all of the relevant non-disjoint geometric models of the input image data. The single entry that features a 'no, blocking ascent' in the fourth column of Table 8 is, accordingly, the only one that is not shaded.

**Table 8:** Numerical values for the ratio of sums of squared residuals of the complex Fourier coefficients of non-disjoint geometric models of the pattern with a large amount of added noise.

| | Left-hand side of (9b) | Right-hand side of (9b) | Inequality fulfilled? |
|---|---|---|---|
| *p2gg* over *p2* | 1.6065574 | 2.04 | yes |
| *p2gg* over *p1g1* | 1.4202899 | 2.0037736 | yes |
| *p2gg* over *p11g* | 1.2564103 | 2.0222222 | yes |
| *c2mm* over *p2* | 1.8852459 | 2.0218182 | yes |
| *c2mm* over *c1m1* | 1.3529412 | 2.0 | yes |
| *c2mm* over *c11m* | 1.5540541 | 2.0 | yes |
| *p4* over *p2* | 1.442623 | 1.9963636 | yes |
| *p4mm* over *p4* | 180.4091 | 1.3333333 | no, blocking ascent |
| *p4gm* over *p4* | 1.2386364 | 1.3454106 | yes |
| *p4gm* over *p2gg* | 1.1122449 | 1.3308081 | yes |
| *p4gm* over *c2mm* | 0.947826 | 1.3370508 | yes |

It is interesting to check if this classification is consistent with the classification of the rather noisy pattern into the most likely projected Laue class as well. Table 9 provides the basis for checking this out. Laue class *4* is, however, still identified by inequality (9b) as the one that minimizes the expected Kullback-Leibler divergence. This could be due to projected Laue class determinations being somewhat less susceptible to added noise, especially to spread noise[A8], than plane symmetry group classifications.

Also, there are many more calculations going into crystallographic symmetry classifications with respect to plane symmetry groups as compared to their counterparts for projected Laue classes. Rounding errors and approximations in the algorithms may therefore accumulate in the calculation for plane symmetry classifications more than for their counterparts for 2D Laue classes.

From the obvious crystallographic inconsistency that plane symmetry group *p4gm* and Laue class *4* have both been identified as K-L best representations of the rather noisy pattern in Figure 6, one needs to conclude that the plane symmetry classification result is incorrect (too high) and Fedorov type pseudosymmetries have been misinterpreted as genuine symmetries. Note that this conclusion is informed by prior knowledge of the crystallographic symmetry classification of the noise-free pattern of the crystal pattern series, but not exclusively based on that knowledge.

**Table 9:** Numerical values for the ratio of the sums of squared Fourier coefficient amplitude residuals of non-disjoint geometric models of the pattern with a large amount of added noise.

|  | Left-hand side of inequality (9b) | Right-hand side of inequality (9b) | Inequality fulfilled? |
|---|---|---|---|
| *4mm* over *4 (in c2mm setting)* | 1.8928571 | 1.333333 | no, but revealing a crystallographic inconsistency |
| *4mm* over *4 (in p2gg setting)* | 1.8214286 | 1.3454106 | no, but revealing a crystallographic inconsistency |
| *4mm* over *2mm (in p2gg setting)* | 1.3076923 | 1.3370508 | yes, as a result of pseudosymmetry |
| *4mm* over *2mm (in c2mm setting)* | 1.2926829 | 1.3246592 | yes, as a result of pseudosymmetry |

Crystallographic symmetry classification results as obtained in this section were to be expected and are in line with those of Moeck and Dempsey (2019) and Dempsey and Moeck (2020) for other series of synthetic crystal patterns with and without added noise that feature pseudosymmetries. The conclusion from all three studies must be that the information theory based classification methods work very well for small to moderate amounts of noise that is to a sufficient approximation Gaussian distributed.

Methods that rely on ignoring higher order terms in equation (3) must, however, fail when there is way too much noise in a more or less 2D periodic pattern that is to be classified with respect to its crystallographic symmetries. Everything depends, of course, also on the relative complexity of a crystal pattern and the strength of its pseudosymmetries.

The identification failure is for the crystal pattern in Figure 6 not "catastrophic" as even when a misidentification is obtained for the most likely underlying plane symmetry group of the noisiest crystal pattern, most human experts would have made the same mistake. Because it is well known that Fedorov type pseudosymmetries are not rare in nature (Chuprunov, 2007, Somov and Chuprunov, 2009), one needs to be extra careful with the crystallographic processing of very noisy images from crystals in order not to misinterpret noise as structural information. Translational pseudosymmetries (de Gelder and Janner, 2005, Somov and Chuprunov, 2009) are also not rare in nature.

In the following subsection, the modestly noisy pattern of Figure 5 is symmetrized to plane symmetry group, *p4*, as this was the Kullback-Leibler best representation of the plane symmetry of that crystal pattern. We will symmetrize the very noisy pattern of Figure 6 to plane symmetry group *p4gm* for demonstration purposes; although our analysis indicated that there was a crystallographic inconsistency, which is to be interpreted as that group being only a pseudosymmetry group.

## 4.4. Results of crystallographic image processing of the two noisy patterns of the analyzed series of crystal patterns

In order to demonstrate the benefits of the crystallographic image processing procedure, the classification results of the noisy patterns in Figure 5 and 6 are now used to boost the signal to noise ratio in these two crystal patterns. Figure 7 shows approximately 2.2 unit cells of the *p4* symmetrized pattern of Figure 5.

The conspicuous bright bow ties in Figure 7 feature site symmetry *2* as perfectly as it is possible for real-world entities that have been derived from disturbed real-world entities by the employed algorithmic crystallographic symmetry enforcing procedure. Note that these bow ties feature point symmetry *2* to a good approximation in Figures 1 to 3 and 5. (This point group represents the highest and second highest site symmetries in plane symmetry groups *p2* and *p4*, respectively.)

Plane symmetry group *p2* was the anchoring group, i.e. the least broken plane symmetry at the $k_l = 2$ or 3 level of Figure 4a. The sum of squared residuals of the complex structure-bearing Fourier coefficients of the *p2* symmetrized model of the crystal pattern in Figure 5 was, accordingly, the lowest in Table 4.

Note how much of the added noise has been removed[A10] by the crystallographic image processing by a visual comparison between the patterns in Figures 5 and 7. This becomes also clear by a comparison of the histogram insets of both figures.

The overall contrast in Figure 7 is lower than in Figure 1. There are also very minor, (almost imperceptible) processing artifacts[A11] in this crystal pattern. These are small prices to pay in the opinion of the author for a significant enhancement of the signal to noise ratio and intrinsic image quality[1] by means of the crystallographic processing of a noisy image. (To see these artifacts more clearly, it might be better to look at the computer screen of the on-line version of this paper in a high magnification rather than directly at a print-out.)

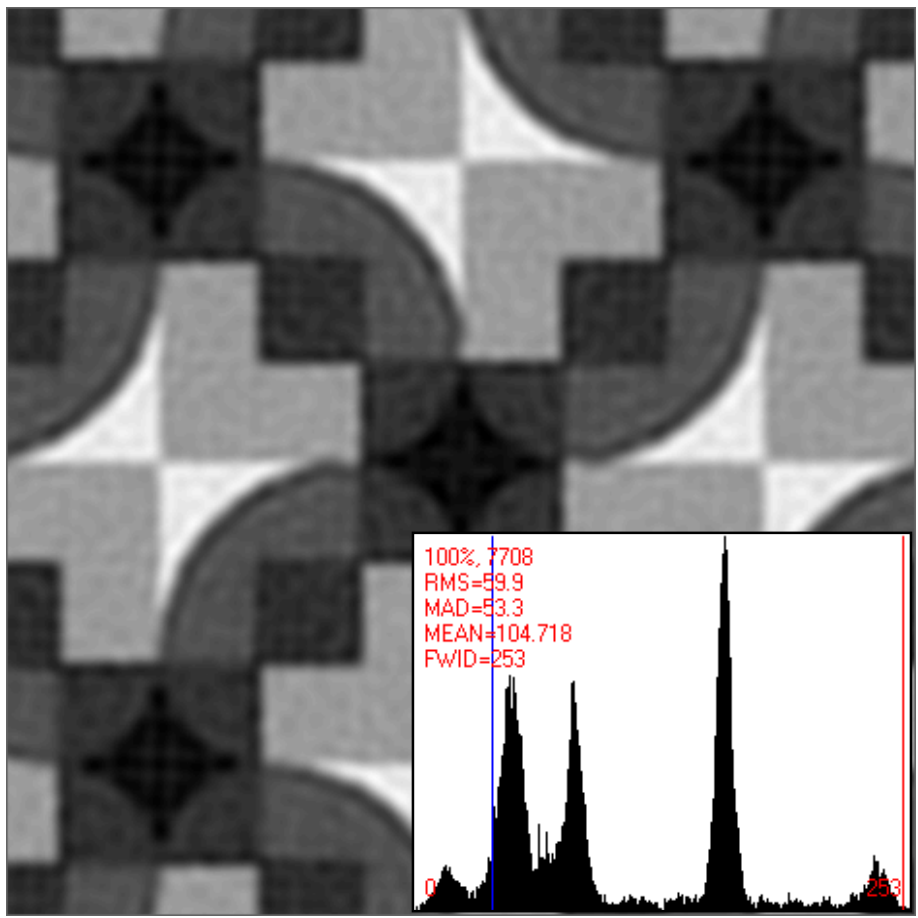

**Figure 7.** Approximately 2.2 primitive unit cells of the moderately noisy pattern of Figure 5 after crystallographic image processing with histogram as inset.

Essentially the same can be said about the crystallographically processed[A10] version of the very noisy pattern in Figure 6. The contrast in the crystallographically processed version of this pattern is in Figure 8 even lower (so that processing artifacts are imperceptible). This is mainly a consequence of using a smaller number of symmetrized complex Fourier coefficients for both the crystallographic symmetry classification and the transformation back into direct space. Note that Figure 8 shows the bright bow ties quite clearly, whereas they were visually unrecognizable (in the absence of prior knowledge) in the crystal pattern that underlies Figure 6.

Because plane symmetry group *p4gm* has been enforced on the very noisy pattern in Figure 6, strong Fedorov type pseudosymmetries have been rendered visibly indistinguishable from genuine symmetries in direct space. The conspicuous bow ties feature in Figure 8, therefore, point symmetry *2mm*, although the corresponding site symmetry in the undisturbed crystal pattern was at best point group *2*, as clearly visible in Figures 2 and 3. Noise in the image has, thus, been misinterpreted as structure as part of a crystallographic image processing that ignored a detected crystallographic inconsistency.

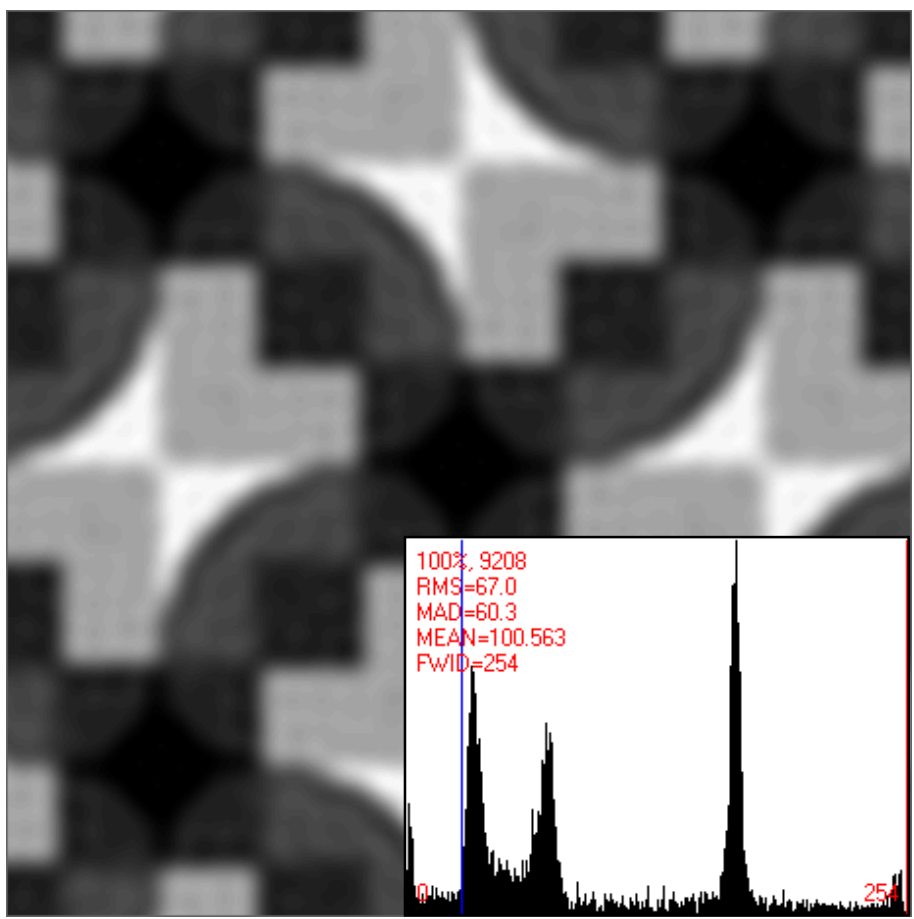

**Figure 8.** Approximately 2.2 primitive unit cells of the rather noisy pattern of Figure 6 after crystallographic image processing with histogram as inset. Note the reduction in contrast and spatial resolution with respect to both the patterns in Figures 1 and 7.

The large amount of added noise pattern, Figure 6, was crystallographically processed in plane symmetry group *p4gm*, Figure 8, for demonstration purposes although the projected Laue class classification, i.e. 2D point group *4*, identified a problem with the *p4gm* classification that is caused by the large amount of added noise. This was done here for the sake of a demonstration of what happens when one symmetrizes a more or less 2D periodic pattern to a plane symmetry group that is not crystallographically consistent with the corresponding 2D Laue class classification by the information theory based methods.

The increased narrowness of the peaks in the histogram inset of Figure 8 with respect to their counterparts in the histogram inset of Figure 7 is due to averaging over twice as many (wrongly identified) asymmetric units during the crystallographic image processing. This wrongful averaging created sites in the translation averaged unit cells that now feature point symmetry group *2mm* at the fractional unit cell coordinates ½,0, 0,½, ½,1, and 1,½, as labeled in Figure 2.

Nevertheless, the suppression of the noise in both of the noisy patterns is quite impressive when judged from the histogram insets in Figures 5 and 6. Again, scanning probe microscopists should take notice of this fact as crystallographic image processing on the basis of objective crystallographic symmetry classifications is now available to them as well. They need, however, to be wary of Fedorov type pseudosymmetries that are easily misinterpreted as genuine symmetries when noise levels are high. Scanning probe microscopists in general and structural biologist who analyze subperiodic intrinsic membrane protein crystals should heed the advice that noisy images are only to be symmetrized to plane symmetry groups that are crystallographically consistent with the projected Laue class classification of a more or less 2D periodic image.

## 5. Comparisons of our classification results to suggestions by the CRISP program and associated comments

The objectively obtained crystallographic symmetry classification results of the previous section are summed up in Table 10 and are now compared to the results of a traditional classification with the electron crystallography program CRISP, Table 11. It is clear from the latter table that the CRISP suggestions do not make distinctions between genuine symmetries and Fedorov type pseudosymmetries.

Note that the comparison of the classification results is based on exactly the same structure-bearing Fourier coefficients and their symmetrized versions as facilitated by using the same *.hka files (without any manual editings[A6]) in both types of classifications for the same pattern area selections.

**Table 10:** Plane symmetry and projected Laue class classifications of the analyzed set of patterns by the author's methods.

| Crystal pattern | Plane symmetry group | Laue class |
|---|---|---|
| Free of added noise, underling Figure 1 | *p4*, with strong *p1g1*, *p11g*, *c1m1*, *c11m*, and somewhat weaker *p2gg*, *c2mm*, *p4gm* Fedorov type pseudosymmetries | *4, 2mm* and *4mm* pseudosymmetries |
| Moderate amount of added noise, underling Figure 5 | *p4*, with strong *p1g1*, *p11g*, *c1m1*, *c11m*, *p2gg*, and somewhat weaker *c2mm*, *p4gm* Fedorov type pseudosymmetries | *4, 2mm* and *4mm* pseudosymmetries |
| Large amount of added noise, underling Figure 6 | *p4*, all Fedorov type pseudosymmetries at the plane symmetry group level were misidentified as genuine symmetries, but the identification of point symmetry *4* as the anchoring Laue class revealed their true nature and confirmed *p4* as the crystallographically consistent plane symmetry group classification | *4, 2mm* and *4mm* pseudosymmetries |

**Table 11:** CRISP program suggestions for the plane symmetry classifications of the analyzed series of patterns.

| Crystal pattern | Plane symmetry group |
|---|---|
| Free of added noise, Figure 1 | *p4gm* |
| Moderate amount of added noise, Figure 5 | *p2gg* |
| Large amount of added noise, Figure 6 | *p4gm* |

As one can interactively test adjacent pattern areas for their CRISP program classification suggestions, one can not only assess the accuracy of that program's classification suggestions but also their precision. It was found that adjacent areas in both the noise-free and moderate amount of noise added pattern resulted in either *p4gm* or *p2gg* classifications with CRISP. The *p4gm* suggestion by CRISP for the noisiest crystal pattern did, however, not change with the selected pattern regions.

At least the noise-free pattern in the series should be homogenous so that all adjacent image areas should be classified as featuring the same plane symmetry. One has to note that a large amount of calculations goes into a plane symmetry classification so that CRISP's symmetry deviation quantifiers for different geometric models of the input image data are indeed slightly different for each different crystal pattern region.

The *p2gg* classification suggestions by CRISP are consistent with the bright bow ties featuring a site symmetry that is no higher than point symmetry group *2*, as clearly revealed in Figures 2 and 3. These classification suggestions assign point symmetry group *2* as well to the centers of the dark curved diamonds in Figure 1, which is a site symmetry underestimation according to the classification results that were obtained with the information-theoretic methods. The strong Fedorov type pseudosymmetries *p1g1* and *p11g* in the selected regions of the noise-free and moderately noisy crystal patterns werey by CRISP misinterpreted as genuine symmetries.

For the modest amount of added noise pattern, see second entry in Table 11, the *p2gg* classification is consistent with the CRISP derived lattice parameter set of $a = 97.1$ pixels, $b = 97.0$ pixels, and $\gamma = 90.0°$ for the crystal pattern in Figure 5. The small difference in the magnitude of the unit vectors should probably be ignored based on what has been shown by Moeck and DeStefano (2018).

Crystallographic symmetry classifications with the CRISP program rely in practice heavily on visual comparisons between the translation averaged (Fourier filtered) and differently symmetrized versions of the input image data by an expert practitioner of electron crystallography. Faced with a *p2gg* classification by CRISP and a 2D Bravais lattice that is almost of the square type (as obtained for the moderate amount of added noise pattern), most electron crystallographers would probably have simply overwritten that suggestion after visual inspections and concluded that the correct plane symmetry group is *p4gm* (based on a square unit cell). In doing so, they would have discounted the possibility of a very strong translational pseudosymmetry or metric specialization (Moeck and DeStefano, 2018).

As mentioned above repeatedly, most human experts would most likely have classified all three synthetic patterns of the series as belonging to plane symmetry group *p4gm* because it would not occur to them that distinctions between genuine symmetries and pseudosymmetries might be necessary. As the analyses in the preceding sections demonstrate, *p4gm* classifications by CRISP for the noise-free and large amount of added noise patterns, see Table 11, constitute overestimations of the plane symmetry that is genuinely there, i.e. *p4*, due to Eva Knoll's handiwork[A1].

Using the author's information theory based methods, no visual comparisons between the translation averaged and differently symmetrized versions of the input image data are necessary. Crystallographic symmetry classifications can, therefore, be made without human supervision, but under the currently necessary assumption that there is indeed more than translation symmetry in a noisy image.

To employ crystallographic image processing techniques, the researcher no longer needs to be an electron crystallographer. This fact allows sufficiently well resolved more or less 2D periodic images from a wide range of crystalline samples that are recorded with different types of microscopes to be processed crystallographically. Previous successes in the crystallographic processing of images from scanning tunneling and atomic force microscopes are quoted by Moeck (2021) and shown in Moeck (2017).

## 6. Summary and Conclusions

Information theory based crystallographic symmetry classification methods for plane symmetry groups and projected Laue classes have been demonstrated on three synthetic crystal patterns. The classifications were for the two noisy patterns complemented by the showing of the corresponding patterns and their histograms before and after their crystallographic processing. Note that these pairs of crystal patterns needed to be shown in this paper for demonstration purposes, but crystallographic image processing by the information-theoretic methods can proceed without prior visual inspections of such patterns by human beings.

It is concluded that the information theory based classification methods are statistically sound and superior to all other existing methods, including the visual insights of human expert classifiers as far as their accuracy at first sight is concerned. Information theory based methods should be developed for crystallographic symmetry classifications and quantifications in three spatial dimensions as there is also subjectivity in the current practice of single crystal X-ray and neutron crystallography[A12]. The detection of noncrystallographic symmetries (defined in the introductory section 1.1 as being incompatible with translation symmetry) is beyond the scope of the demonstrated methods and there are no plans by this author to try to tackle that kind of problem.

## Notes added in proof

**1.** As quoted in Moeck (2018 and 2019), there is a *direct space* G-AIC approach by Xanxi Liu and coworkers to plane symmetry group classifications of more or less 2D periodic patterns. The number of analyzed translation periodic tiles, *t*, in the crystal pattern enters in that approach the direct-space analog to (9a) so that

$$\frac{\widehat{J}_m^{\;direct}}{\widehat{J}_l^{\;direct}} < 1 + \frac{2(k_m - k_l)}{k_m(t \cdot k_l - 1)} \tag{9c}$$

results. There is no translation averaged unit cell and with that no *p1*-symmetrized model of the input image data in that approach, so that the benefits of substantial noise reductions by working exclusively with the periodic structure-bearing Fourier coefficients vanish. For *t* > 1, a non-zero ratio of sums of squared *direct space* pixel-intensity residuals for ascent to a geometric model of the data with $k_m = 2$ or 3 is, however, defined by (9c). (This might be the only[A10] advantage of working in direct space.) When all of the sums of squared complex Fourier coefficient residuals (equation 1) at the $k_m = 2$ or 3 level of the plane symmetry hierarchy tree (Fig. 4a) are rather high, using inequality (9c) with $k_m = 2$ or 3, $k_l = 1$, and *t* > 1 could either help with the identification of the anchoring plane symmetry group or provide a statistical proof that there is only translation symmetry and in the crystal pattern. This would, however, work reliably only for low and moderate levels of approximately Gaussian distributed noise. The propensity of misidentifying Fedorov type pseudosymmetries as genuine symmetries increases in a direct space approach more strongly with the noise level than in the present study (that was performed exclusively in Fourier space).

**2.** The development of an information theory based method for the classification and quantification of electron diffraction patterns, as motivated at the end of the third appendix (C.2), progresses well. The first objective projected point symmetry classifications and quantifications results were obtained from an experimental spot pattern, as discussed in Moeck and von Koch (2022).

**Acknowledgements**   The current fellow members of Portland State University's Nano-Crystallography Group, Regan Garner, Choomno Moos, Grayson Kolar, Gabriel Eng, Noah Allen, and Lukas von Koch are thanked for critical proofreads of the manuscript. Regan Garner is also thanked for the graphs in Figures 4a and 4b. Professor Eva Knoll of the Department of Mathematics of the University of Quebec at Montreal is thanked for both a critical proofread and

## Appendices:

### A. Notes on the text

**A1.** The artist Eva Knoll painted a single asymmetric unit onto a single ceramic tile by hand, see the last appendix in arXiv: 2108.00829. (That reference is to a significantly expanded version of this paper where the artist describes the genesis of "Tiles with quasi-ellipses" in her own words and gives a reference to her portfolio). The painted asymmetric unit featured a broken mirror line across one of its two diagonals, but covered the whole ceramic tile. That tile had a square shape (to a very good approximation) and was 6 inches (15.24 cm) long on its edges. For a color reproduction of the original painted tile, see arXiv: 2108.00829.

The artist took a color photo of that square and produced multiple copies of that photo with the shape of squares of the same size. Sets of four photos of the tile were assembled into four-fold larger squares with four-fold rotation points at their centers by making sure that the broken mirror lines run along the fractional coordinates x,x+½, -x,-x+½, -x+½,x, and x+½, -x of the thus created unit cell. (The multiplicity of the general position in this primitive unit cell is four.) It is quite remarkable that three pairs of slightly broken glide lines were created in the unit cell as a result of this assembly process.

The so created (four-fold larger) unit cell squares were then laid out on a square Bravais lattice without overlaps or gaps. This created four-fold rotation points at each of the four vertices of the unit cell and two-fold rotation points in the middle of each of its four edges.

The whole piece of Eva Knoll's graphic artwork consists, thus, of a translation periodic array of four properly assembled photocopies of her original tile (asymmetric unit). The graphic artwork features plane symmetry group *p4* as the result of its creation process. (The genuine site symmetries in the assembly are point groups *4* and *2*, which are non-disjoint.)

The artistically sophisticated distribution of paint, the broken mirror line in the original asymmetric unit, and the two- and four-fold rotation points that resulted from the translation-periodic assembly process combined to several Fedorov type pseudosymmetries. The latter give the visual impression that the graphic artwork features a unit cell with plane symmetry group *p4gm*, at least at first sight.

Due to the large reduction in the size of the photocopies of the original tile, the diagonal pseudo-mirror line of the original tile feigns a genuine mirror line pretty well, at least at first sight. The gray scale reproduction of the original digital-color artwork in Knoll (2003) has an edge length of 5.7 cm only (and is of a square shape). There was, thus, a linear reduction of the edge length of the original painted tile to one of its digital photocopy counterparts by approximately a factor of 21.

The artist also created random assemblies of photocopies of her original tile without gaps or overlaps, see the last appendix in arXiv: 2108.00829 for a color version of such an assembly.

**A2.** So far unpublished results on the classification of parallel-illumination transmission electron microscope images from a subperiodic intrinsic membrane protein crystal are mentioned in appendix C briefly. The ongoing development of an information theoretic classification and quantification method for projected crystallographic point symmetries from transmission electron diffraction patterns in approximate zone axis orientations is also mentioned in appendix C.

That method has the potential to (*i*) distinguish genuine quaternary symmetries of intrinsic membrane protein complexes from pseudosymmetries at the point symmetry level and (*ii*) solve the symmetry inclusion problem in a recently demonstrated symmetry-contrast mode (Krajnak and Etheridge, 2020) of 2D scanning transmission electron microscopy on a 2D grid with fast pixelated direct electron detectors (Ophus, 2019, commonly referred to as 4D-STEM).

**A3.** The obtaining of satisfactory Fourier filtering results was facilitated by the above-mentioned increase in the number of unit cells in the crystal pattern that underlies Figure 1 by computational periodic motif stitching. This kind of computational increase of a digital image of the original graphic work of art is also highly beneficial to the subsequent crystallographic symmetry classification and a possible follow up step of the enforcing of the plane symmetry that most likely underlies the pattern in a statistically sound sense.

Note also that Fourier filtering (Park and Quate, 1987) is an integral part of symmetry classifications and any subsequent crystallographic processing of a digital image. This is because the sums of squared residuals and the symmetrizing of the input image data are based only on the structure-bearing Fourier coefficients of a digital image (that are laid out on a lattice in reciprocal space).

**A4.** The analogy between Wyckoff positions in the direct space unit cell of an ideal crystal pattern and so called "domain maps" (Verberck, 2012) of the symmetries of the Fourier coefficients of such a pattern may be helpful to appreciate this statement. There is also an analogy between the asymmetric unit in direct space and the "minimal domain" in Fourier space.

Typically, there are many more general Wyckoff positions with site symmetry *1* and their characteristic multiplicity than special Wyckoff positions with higher site symmetries and their reduced multiplicities. For unit cells that contain a large number of points in direct space, the multiplicity of the general Wyckoff position approximates the combined-weighted multiplicities of all Wyckoff positions in an ideal crystal pattern of high complexity reasonably well.

**A5.** This is because the dimension of the data space is in our case one, i.e. intensity values of pixels. The co-dimension is the difference between the dimension of the data space and the dimension of the model space, $d$ in (3) and (4). The dimension of the model space is zero, in our case, as geometric points are representations of the individual pixels.

**A6.** Relying on the *.hka files of CRISP without further editing is not ideal, see also note A7, but was done in this study in order to enable a direct comparison of the symmetry classification results. The geometric models that are represented by *.hka files with different numbers of data points, different dynamic ranges, and different spatial resolutions do not necessarily always give the best possible symmetrized version of the input image data in Fourier space. For the purpose of the demonstrations in this paper and to allow for the comparison of classification results that were obtained with the information theory based methods to those of the CRISP program, the accuracy of all geometric models is deemed to be more than sufficient.

On all accounts, the geometric models that CRISP provides in the form of exportable *.hka files are always quite representative of symmetrized versions of analyzed images as demonstrated by the successes of countless electron crystallography studies despite necessarily different choices for the dynamic range, spatial resolution, and numbers of included structure-bearing Fourier coefficients.

**A7.** Ideally, one would base all calculations on symmetrized models of the input image data that feature exactly the same appropriately indexed structure-bearing Fourier coefficients and number of such coefficients. To obtain the same number of data points (complex Fourier coefficients of the image intensity) in all geometric models of the input image data, one would need to treat Fourier coefficients that are absent in certain geometric models as featuring zero amplitude and arbitrary phase. The absences can either be systematic or incidental. In both cases, the zero amplitude Fourier coefficients are characteristics of the properly symmetrized geometric models of the input image data.

One can then give confidence levels for the classification into minimal supergroups over maximal subgroups by using equations (B-1) to (B-4) of the second appendix and provide a complete crystallographic symmetry measurement result. In the absence of generalized noise (including small calculation errors), the smallest possible entry in the second column of Table 2 should for genuine symmetries then be restricted to unity.

**A8.** Spread noise "mimics" to some extent the effects of small random crystal-sample movements in a microscope during the recording of a more or less 2D periodic image.

**A9.** The fat tails in the histogram in Figure 6 are actually an artifact of the way the Gimp program (Program I) adds Gaussian distributed noise to the individual pixel intensity values. All pixel intensities that would after the adding of the noise be below zero are set to zero (black) and all pixel intensities that would be larger than 255 are set to 255 (maximal brightness). This fat-tails effect can also be seen in the histogram of the moderately noisy crystal pattern that underlies Figure 5.

The histogram in Figure 6 may to a better approximation be described by one of Mandelbrot's stable distributions (Mandelbrot, 1963). Such a distribution may acquire approximate Gaussian tails with the addition of more stably-distributed noise from a multitude of sources. This is in line with Mandelbrot's bon mot: *"approximations are absurd in some problems but are adequate in many others, and they are so simple that one must consider them first"*, 1963. The central limit theorem applies to both stable distributions and Gaussian distributions.

**A10.** Note that much of the noise removal is due to the translation averaging by Fourier filtering over approximately 88 unit cells. In order to obtain a good image-quality enhancement in an experimental study of a crystal, one needs to start with an image with a large field of view and medium magnification. That is somewhat unusual in the microscopical practice where the focus is often on structural defects and images are recorded with small fields of view and very high magnifications.

As discussed in detail in Moeck, 2019, the Fourier space approach to crystallographic symmetry classifications and the subsequent optimal processing of a 2D crystal pattern offers significant advantages over any direct space approach. Wiener filters can be used in *direct space* to increase the image quality, but that does not restore the broken site symmetries in the translation averaged unit cell.

The precondition for using the Fourier space approach is, on the other hand, a direct space image with a sufficient number of more or less translation periodic unit cells which are represented by a large number of pixels. Depending on the complexity of the unit cell, several tens of unit cells may suffice for good image processing results. The results of the processing of larger regions of more or less 2D periodic images are always better than their counterparts for smaller regions (Dempsey and Moeck, 2020). As for the shape of the processed image regions, circular disks are preferable over any other shapes. In the electron crystallography of intrinsic membrane protein crystals, one typically averages over several hundreds to a few thousands unit cells and uses magnifications of around 50,000 only.

The averaging of the periodic structure-bearing Fourier components with matching Laue indices from multiple images of the same crystalline sample and plane symmetry group *p1* is analogous to merging X-ray or neutron diffraction data from several crystals of the same kind and common practice in electron crystallography.

The stitching together of experimental direct space images that were recorded under different imaging conditions in order to increase the number of unit cells in the composite image is not recommended. Using a computer program such as Microsoft ICE 2.0 (Program II), this may lead to additional Fourier coefficients that represent the created superstructure. The stitching together of the crystal pattern that underlies Figure 1 did not lead to additional Fourier coefficients because it was free of noise, i.e. all unit cells were exactly identical due to the creation process of the graphic piece of art, see note A1.

**A11.** Note that the "faint square crosses" inside the "dark curved diamonds" with site symmetry *4* in Figure 7 at the unit cell coordinates 0,0, 1,0, 01, and 1,1 (as marked in Fig. 2) originate partly from the tiling of digital photos of the same painted ceramic square tile, see Fig. A-7 in the expanded on-line version of this paper (arXiv: 2108.00829). There are corresponding "narrow cross" features at these positions in the expanded digital version of Eva Knoll's piece of graphic art that served as basis of all demonstrations in this paper and is available in the *.jpg and *.tif format in the supporting material of this paper.

The very low-contrast "four-fold feature" inside the dark curved diamond at the fractional unit cell coordinate ½,½ originate mainly from the "symmetrization of remains of the added noise" by the crystallographic image processing. Analogously, note that the bright bow ties (at fractional unit cell coordinates ½,0, ½,1, 0,½, and 1,½, as marked in Fig. 2) are not homogenously bright (as they appear to be in Figures 1 and 8). They feature instead a "fine structure" with the intensity distribution of a two-fold rotation point and originate partly from the symmetrization of local Fourier ripples. A more thorough discussion of these artifacts is provided in arXiv: 2108.00829.

All of these artifacts could have been suppressed by larger spatial and dynamic range restrictions of the noisy structure-bearing Fourier coefficients in Fourier space, resulting unavoidably in lower contrasts in the direct space pattern after back transforming. This has in principle been demonstrated with the processing of the nosiest crystal pattern of the series, see Figure 8.

**A12.** In every single crystal X-ray or neutron diffraction based determination of an unknown crystal structure, one needs to assign a space group in which the subjectively most reasonable model for the structure is to be refined. Information theory, as defined in footnote 2, is partly about the selection of the model for experimental data that is statistically/objectively most justified by the data itself. Since the experimental data is in diffraction based crystallography of a geometric nature, a geometric form of information theory such as the one by Kenichi Kanatani is applicable.

When the symmetry classification (and quantification) methods of this paper have been generalized to three spatial dimensions, Walter C. Hamilton's well known significance tests of crystallographic R-values after refinements into non-disjoint space groups (Hamilton, 1965) could be considered superseded. This is because they have been set up as null-hypothesis tests. Information theory is widely considered to offer a superior alternative to null-hypothesis testing, see Anderson, 2008, for a gentle introduction on how to bring more objectivity to scientific studies.

**B. Ad-hoc defined confidence levels for classifications into minimal supergroups for a special case of the inequality on which the author's information theory based methods are based**

For the special case $N_m = N_l$, inequality (9b) reduces to

$$\frac{\bar{J}_m}{\bar{J}_l} < 1 + \frac{2(k_m - k_l)}{k_m(k_l - 1)}$$ , which has been labeled as inequality (9a) in the main part of this paper.

When $N_m = N_l$, one can take advantage of the inequality having the simple form of a numerical value on its right-hand side that is just the sum of unity and a constant term that only depends on the difference in the hierarchy levels, $k$, of the respective two symmetrized non-disjoint models that are to be compared to each other, see Figures 4a and 4b. The respective ratios of sums of squared complex Fourier coefficient residuals and sums of squared Fourier coefficients amplitudes are provided in these figures as insets for easy reference. (The comparison of two non-disjoint symmetrized models with respect of their ability to represent the input image data is based on having an appropriate "relative measure" of their numerical distance to the common translation-averaged-only model in the first place.)

Inequality (9a) can be used in connection with ad-hoc defined confidence levels for geometric model selections. Providing such confidence levels can be understood as giving a quantitative measure of the corresponding model-selection uncertainty, which needs to accompany any crystallographic symmetry measurement results in order to be complete (Helliwell, 2021).

Based on Kanatani's information content ratio equation (Kanatani, 1998), ad-hoc defined confidence levels for model selections in favor of a non-disjoint more symmetric/restricted geometric model can for the special case $N_m = N_l$ be straightforwardly defined (whenever inequality (9a) is fulfilled). For two non-disjoint geometric models one obtains:

$$K = \sqrt{\frac{1 - \frac{1}{k_l}}{1 + \frac{1}{k_l}}\left(\frac{\hat{J}_m}{\hat{J}_l} + \frac{\frac{2}{k_m}}{1 - \frac{1}{k_l}}\right)} \leq 1$$

(B-1),

and the critical value for $K$ is obtained for inserting the condition

$$\frac{\hat{J}_m}{\hat{J}_l} = 1$$

(B-2),

into (B-1) so that

$$K_{critical} = \sqrt{\frac{k_m - \frac{k_m}{k_l} + 2}{k_m + \frac{k_m}{k_l}}} < 1$$

(B-3)

results.

Obviously, $K \geq K_{critical}$ is valid as the ratio of the two sums of squared residuals ranges from unity (B-2) to a constant value that is larger than unity and depends on the particular combination of $k_m$ and $k_l$ in inequality (9a).

When the ratio of the squared residuals is unity (as in (B-2)), one has 100 % confidence in choosing the more symmetric model over the less symmetric model. Both models fit the input image data equally well in that special case, which will in practice only be obtained for noise-free mathematical idealizations of real world images, perfect geometric models, and with a perfectly accurate algorithm. When inequality (9a) is not fulfilled, one has zero confidence in the selection of the more symmetric model over its less symmetric counterpart. This is all formalized by the definition of the confidence level in identifying a minimal supergroup over its maximal subgroup

$$C_m = \frac{1 - K}{1 - K_{critical}}(100\%)$$

(B-4),

which takes on values between 100 % and zero as a function of the ratio of the sums of squared residuals. (Negative values, which are meaningless, result from (B-4) when inequality (9a) is not fulfilled so that K > 1.) It makes sense to define an average confidence level for a transition from all maximal subgroups to their common minimal supergroup. For small symmetry breakings of each individual maximal subgroup or class and low-noise data, this average confidence level can be rather high.

## C. Outlooks on ongoing developments of the information theoretic crystallographic symmetry classification and quantification methodology and their potential applications

Formulations of geometric information criteria are possible where the generalized noise does not need to be approximately Gaussian distributed. For a non-Gaussian noise model, the appropriate logarithmic likelihood estimate needs to be used instead of a sum of squared residuals. The generalized inverse of the Fisher information matrix needs then to replace the isotropic covariance matrix of Gaussian distributed noise. In Kenichi Kanatani's own words: *"such an extension does not seem to have much practical significance because of the difficulty of estimating the parameters of a non-Gaussian noise distribution"* (1998). Note that the generalized noise arises from multiple sources with different characteristics, but the overall distribution is not supposed to be dominated by anyone of these sources.

The assumption had to be made in the main part of this paper that there is indeed more than translation symmetry in a more or less 2D periodic pattern that is to be classified with respect to its crystallographic symmetries. This may, however, not always be the case.

There are certainly approximately 2D periodic patterns with and without noise in which all point/site symmetries higher than the identity operation are only pseudosymmetries and not genuine. These patterns are revealed by large sums of squared complex Fourier coefficient residuals for all plane symmetry groups with $k_l = 2$ and 3 and large sums of squared Fourier coefficient amplitude residuals for all projected Laue classes with $k_l = 4$ and 6 (Note that the definition of crystal pattern (Dictionary III) leaves it open if there are site/point symmetries higher than the identity operation or a single glide line in the pattern or not.)

Those crystal patterns or images of crystal structures would be misclassified by the author's methods at the present stage of their development if the facts were ignored that the sums of squared residuals for all of these groups and classes are rather large. The first of the "notes added in proof" above identifies a practical workaround to this problem.

## C.1. Quaternary symmetry and pseudosymmetry of intrinsic membrane protein complexes

Crystallographic studies of the quaternary structure of intrinsic membrane protein complexes in lipid bilayers are in the structural biology field based on parallel-illumination transmission electron microscope (TEM) images that are dominated by Poisson distributed shot noise. As mentioned at the beginning of the previous subsection, an information theoretic approach to the classification and quantification of crystallographic symmetries in such highly beam-sensitive crystalline samples (and the digital images that were recorded from them) could be specifically developed by a generalization of Kanatani's geometric framework.

For the time being, this author sees no harm in using the methods of this paper in that particular field as well. This is for two reasons: (*i*) because shot noise becomes with moderate electron doses approximately Gaussian distributed and (*ii*) the subjective (and less accurate) traditional crystallographic symmetry classification methods (that do not model the noise at all) are currently used for exactly this purpose.

So far unpublished results of this author on the plane symmetry group and Laue class classification of the cyclic nucleotide-modulated potassium channel MloK1 from bacterium mesorhizobium loti in both the open and closed conformations indicate that the projected genuine, i.e. least broken, quaternary symmetry of this protein complex is point group *2*. There is, however, a strong four-fold pseudosymmetry along the channel axis as indicated by the relatively low sum of squared residuals of the complex Fourier coefficients for plane symmetry group *p4gm*.

This makes the potassium channel a dimer of two dimers, while other authors (Chiu and co-workers, 2007, Kowal and co-workers, 2014 and 2018) claimed it to be a tetramer. Their claim relies, however, on the traditional crystallographic symmetry classification methodology, which contains elements of subjectivity.

Incidentally, the experimental facts of this author's study on the above mentioned MloK1 potassium channel are similar to the results of the information-theoretic analysis of the noisiest crystal pattern in the main part of this paper. The histograms of the experimental TEM images revealed a single broad peak with slim tails and a mean value that corresponded to approximately 50 % of the whole dynamic intensity range. This peak looked visually like some Gaussian function to a much better approximation than the histogram inset in Figure 6. In other words, there was apparently enough shot noise in the experimental images and contributions from other noise sources so that the generalized noise became approximately Gaussian distributed.

According to other authors (Chiu and co-workers, 2007, Kowal and co-workers, 2014 and 2018), the plane projected symmetry of MloK1 potassium channel crystals from this bacterium in lipid bilayers is plane symmetry group *p4gm*. This author's analysis indicates, on the other hand, that this can only be a strong pseudosymmetry because projected Laue class *2mm* has been identified as the K-L best representation of the symmetry information in the amplitude maps of the discrete Fourier transforms of the TEM images. Note that this analysis was based on some of the same experimental images that Kowal and co-workers (2014) used in their study, as downloaded from the EMDataResource (Database in list of references). Those experimental images were recorded by these other authors with a large underfocus at a nominal zero-tilt setting of the specimen goniometer. A tomographic images and derived electron density maps supported model mechanism for the opening and closing of this particular potassium channel that is restricted to four-fold rotation symmetry, as the one proposed by Kowal and co-workers (2018) has, accordingly (at the present time) less "geometric support" than an alternative mechanism that is restricted to two-fold rotation symmetry only.

Note that the identification of projected Laue class *2mm* as point symmetry of the K-L best model of the experimental data rules out the existence of genuine four-fold rotation points as site symmetries in the unit cell of the MloK1 potassium channel crystal in an analogous manner, as the entries for projected Laue class *4* in Tables 7 and 9 rule out plane symmetry group *p4gm* for the very noisy crystal pattern in Figure 6. It is notable that it was again the information theoretic projected Laue class determination that led to the identification of a strong Fedorov type pseudosymmetry at the site/point symmetry level. Presumably, projected Laue class determinations by the new method are less sensitive to noise than the corresponding plane symmetry group determinations. (Amplitude maps of discrete Fourier transforms of perfect crystal patterns are known to be translation invariant.)

Complementing information-theoretic classification studies of transmission electron diffraction spot patterns from intrinsic membrane protein complexes under zero-crystal-tilt condition would be helpful as these patterns typically feature more spots than the number of structure bearing Fourier coefficients of the corresponding TEM images and the spot intensities are not affected by aberrations of the objective lens. This means they contain more point/site symmetry specific information. Electron diffraction patterns from perfect plane-parallel crystals are translation invariant in an ideal TEM so that small random sample movements under the electron beam might be tolerable when projected point symmetry classifications are made on the basis of such patterns.

**C.2. Development of an information-theoretic projected point symmetry classification and quantifications method**

A first motivation for the development of an information-theoretic projected point symmetry classification and quantifications method was provided in the last paragraph of the previous sub-section. There are, in addition, very interesting developments in 4D-STEM (Opus, 2019) with fast pixelated direct electron detectors. A new symmetry-contrast imaging mode has, for example, been recently demonstrated by Krajnak and Etheridge (2020).

Future developments of that contrast mechanism into a widely accepted standard are, however, hampered by the well known symmetry inclusion relationships. The incorporation of a newly developed information-theoretic projected point symmetry group classification and quantification method from experimental electron diffraction patterns would solve this problem. As $N$ is not likely to be large in electron diffraction patterns of crystals with small unit cells, suitable replacements for equation (6) have to be used.

This author has taken up the challenge to develop such a method for selected area electron diffraction spot patterns, precession electron diffraction patterns, nearly parallel-illumination nano-diffraction disk patterns, and convergent beam micro-diffraction patterns with essentially non-overlapping and featureless (blank) electron diffraction disks.